\documentclass[twocolumn, superscriptaddress, nofootinbib]{revtex4-2}

\usepackage{graphicx}
\usepackage{xcolor}
\usepackage{amsmath}
\usepackage{amssymb}
\usepackage{csquotes}
\usepackage{soul}

\begin{document}

	\title{A decoder for the triangular color code by matching on a M\"obius strip}

	\author{Kaavya Sahay}
	\affiliation{Department of Physics, Indian Institute of Technology Delhi, Hauz Khas, New Delhi, 110 016, India}
	
	\author{Benjamin J. Brown}
	\affiliation{Centre for Engineered Quantum Systems, School of Physics, University of Sydney, Sydney, New South Wales 2006, Australia}
	\date{\today}
	
	\begin{abstract}
		The color code is remarkable for its ability to perform fault-tolerant logic gates. This motivates the design of practical decoders that minimise the resource cost of color-code quantum computation. Here we propose a decoder for the planar color code with a triangular boundary where we match syndrome defects on a nontrivial manifold that has the topology of a M\"{o}bius strip. A basic implementation of our decoder used on the color code with hexagonal lattice geometry demonstrates a logical failure rate that is competitive with the optimal performance of the surface code, $\sim p^{\alpha \sqrt{n}}$, with $\alpha  \approx 6 / 7 \sqrt{3} \approx 0.5$, error rate $p$, and $n$ the code length. Furthermore,  by exhaustively testing over five billion error configurations, we find that a modification of our decoder that manually compares inequivalent recovery operators can correct all errors of weight $\le (d-1) /2$ for codes with distance $d \le 13$. Our decoder is derived using relations among the stabilizers that preserve global conservation laws at the lattice boundary. We present generalisations of our method to depolarising noise and fault-tolerant error correction, as well as to Majorana surface codes, higher-dimensional color codes and single-shot error correction.
	\end{abstract}

	\maketitle

	\section{Introduction}
	A quantum computer must be able to perform information-processing tasks with near noiseless logical qubits. To deal with the noise that physical qubits will experience, we imagine protecting and processing quantum information using quantum error-correcting codes~\cite{Shor95, Shor96, Calderbank96, Steane96, Kitaev03, Dennis02, Terhal15, Brown16, Campbell17}. As such we seek codes that can perform logical operations efficiently, while dealing with the significant number of errors that physical qubits will suffer. Ideally, we will find resource efficient codes whose construction will require a relatively small number of the qubits that are currently available in laboratories, that also respect the technical constraints that are imposed by modern hardware~\cite{Kelly15, Corcoles15, Takita16}.

	Color codes~\cite{Bombin06, Bombin07a, Bombin13, Bombin15} can be realised with a planar layout~\cite{Bombin06, Bombin18} and, moreover, require a relatively small number of qubits to encode a logical qubit at some designated code distance~\cite{Bombin07resources, Fowler11, Landahl14, Lavasani18, Kesselring18}. Ultimately, better decoders will reduce the resource cost of fault-tolerant quantum computation using the low-overhead logical operations that are permitted by color codes~\cite{Bombin06, Bombin07a,  Fowler11, Landahl14, Bombin15, Bombin14b, Kesselring18, Chamberland20, Beverland21}.

	We aim to reach a very low logical failure rate with a minimal number of physical qubits~\cite{Dennis02, Bravyi13, Fowler12a, Watson12, Beverland19}. At low error rates, and neglecting entropic factors, we expect the logical failure rate to decay like ${P}_{fail}\sim p^{t}$ where $p$ is the error rate of the physical qubits and $t \le (d-1) / 2$ is the number of errors the code can tolerate with $d$ the code distance. Maximising $t$ will optimise the performance of the code far below threshold.

	In addition to finding high-performance decoders it is also important for them to be practical. That is, they should have a fast runtime and they should be versatile to realistic laboratory settings. To this end we turn to the minimum-weight perfect-matching algorithm~\cite{Edmonds65, Dennis02, Kolmogorov09,Higgott21}. Decoders based on matching generalise naturally to the fault-tolerant setting where stabilizer measurements are unreliable~\cite{Dennis02, Wang03, Raussendorf07, Brown20parallelized}. Moreover, the matching subroutines can be replaced with almost linear-time algorithms that demonstrate comparable performance~\cite{Delfosse17, Huang20}.

Here we propose an efficient matching decoder for the color code with boundaries that corrects high-weight errors. We find that a basic implementation of our decoder on the hexagonal lattice demonstrates a logical failure rate competitive with the square-lattice surface code~\cite{Dennis02, Bravyi13,  Watson12, Beverland19} at low error rates using an equivalent number of qubits. We report a logical failure rate that scales as $\sim p^{t}$ with $ t \approx 0.42 d \approx 3 d / 7 $ using an independent and identically distributed noise model. Given this instance of the color code requires $n = O(3d^2 / 4) $ physical qubits for its realisation, we have that $t \approx  6  \sqrt{n}  / 7\sqrt{3} \approx 0.5 \sqrt{n}$ which is competitive with the optimal performance of the surface code at low $p$, that demonstrates $ t = \sqrt{n} / 2$~\cite{Dennis02, Bravyi13,  Watson12, Beverland19}. Our decoder also demonstrates a threshold $p_c \sim 9.0\%$, exceeding that of other matching decoders on the hexagonal lattice.

Of course, we should strive to find decoders that maximise the number of errors a code can tolerate with $t = (d-1) / 2$. We improve our decoder by developing a method for obtaining two inequivalent low-weight corrections, see also~\cite{Hutter14}. This method enables us to manually compare different corrections returned from the matching subroutines to make a better choice of output. We find that our improved decoder corrects errors up to its distance for system sizes for $d \le 13$. We obtain this result by exhaustively testing all errors of weight $\le (d-1)/2$ for each system size. At $d = 13$ we check over five billion error configurations.

	A number of decoders have been developed for the color code and its variants~\cite{Landahl11, Bombin12, Sarvepalli12, Brown16a, Delfosse17, Aloshious18, Li18, Tuckett19, Turner20, Li20, Chubb21, Sabo21} including several matching-based decoders~\cite{Wang10,  Delfosse14, Stephens14, Delfosse17, Aloshious18, Li18, Kubica19, Chamberland20, Beverland21}. Broadly speaking, two different schools of thought have led to matching decoders. The first, unfolding~\cite{Bombin12, Bhagoji15, Kubica15, Zhu20}, takes a physical perspective~\cite{Chen10} to find a unitary operator that maps the color code onto two copies of the surface code. The other, projection~\cite{Delfosse14}, reduces the dimensionality of the objects of the color code lattice using tools from homology. Both approaches map the color code onto copies of the surface code and a correction is found in the latter picture.

	Fundamentally, the surface code permits the use of matching decoders due to its materialized symmetries~\cite{Kitaev03,Brown20parallelized} where relations among the local elements of the stabilizer group give rise to a defect parity conservation law. Given that errors always produce defects in pairs, we can locally match nearby defects to successfully correct the code with high probability. In~\cite{Brown20parallelized} it was proposed that the symmetries of more general stabilizer codes offer a unifying picture to find matching decoders for other codes, see e.g.~\cite{Tuckett20, Bonilla-Ataides21, Nixon21}. At a basic level, we find that this perspective reproduces the aforementioned strategies of decoding the color code. Further, a more careful examination of the symmetries at the boundary of the color code allows us to find a matching graph that is associated to a global symmetry that is embedded on a M\"{o}bius strip. It is this observation that enables us to produce our results.

To elaborate on some of the principles we use to derive our matching decoder, in addition to our numerical results, we also give an extended discussion on how the ideas we have used can be generalised for other decoding problems with the color code and its variants. We look at color codes with different boundary conditions, Majorana surface codes~\cite{Vijay15, Litinski18, Li18, Viyuela19, Chao20}, higher-dimensional color codes~\cite{Bombin13, Bombin15, Kubica15a}, and we discuss single-shot error correction with the gauge color code~\cite{Bombin14a, Brown16, Beverland21}. We also examine the depolarising noise model together with other types of unfolding~\cite{Kesselring18, Brown20parallelized}, as well as fault-tolerant error correction~\cite{Stephens14, Chamberland20, Beverland21}.
	
	In what follows, we briefly introduce the color code and describe our decoder from the perspective of symmetries in Sec.~\ref{Sec:Background}. In Sec.~\ref{Sec:Decoder} we argue that our decoder will be capable of decoding high-weight errors. We evaluate the performance of our decoder in Sec.~\ref{Sec:Results} using several numerical experiments before offering some concluding remarks. We go into further detail about the matching subroutines used by our decoder in Appendices~\ref{App:SymmetriesAndLogicals}, \ref{App:MWPM},~\ref{App:MeasuringEdgeWeights} and~\ref{App:AlternativeCorrections}, and how we analyse our data in Appendix~\ref{App:Fitting}. We give an extended discussion on the generalisations of our decoding methods in Appendix~\ref{Sec:Generalisations}.
	
	\section{The color code}
	\label{Sec:Background}
\begin{figure}
		\includegraphics{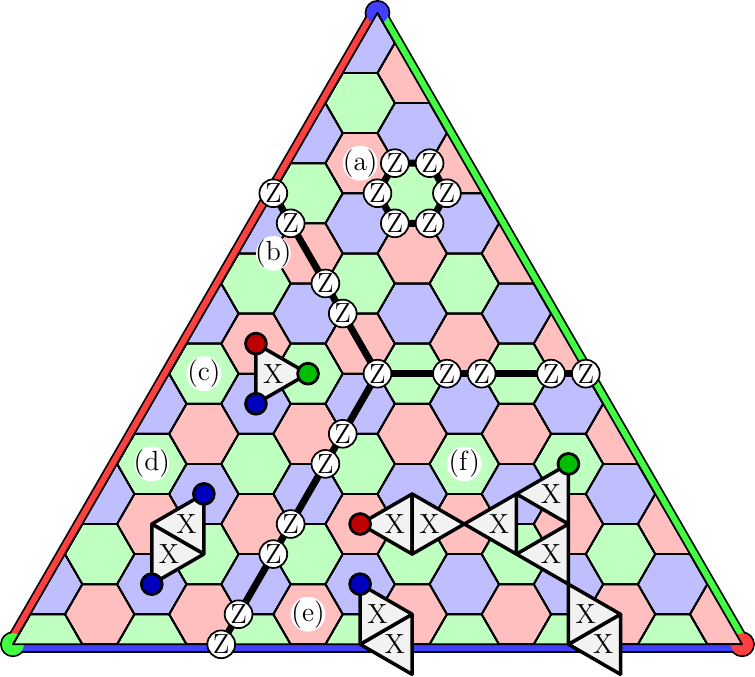}
		\caption{\label{Fig:ColorCode}The color code on the lattice with triangular boundaries. Qubits lie on the vertices of the three-colorable lattice and the red, green and blue boundaries of the lattice are shown by brightly colored lines of their corresponding color. (a)~A stabilizer generator $ S_f$. (b)~A string-like logical operator terminates at each of the three distinct boundaries of the lattice and branches in the middle of the lattice. Examples of bit-flip errors and the individual defects of their syndrome are shown at~(c),~(d),~(e) and~(f).}
	\end{figure}

	We define the color code~\cite{Bombin06} on a two-dimensional lattice with three-colorable faces. That is the faces, indexed $f$, can be assigned one of three colors, red, green and blue, such that no two faces of the same color are touching. We will focus on the hexagonal lattice shown in Fig.~\ref{Fig:ColorCode} for simulations but we remark that the discussions we give are agnostic of the underlying geometry of the three-colorable lattice. Let us label the colors with boldface symbols from the set $\mathcal{C} = \{ \mathbf{r},\, \mathbf{g},\, \mathbf{b} \}$, and we define the function $\mathbf{col} : o \rightarrow \mathcal{C}$ that specifies the color of some object of the lattice $o$. It will also be helpful to assign a color to each of the edges of the lattice. We say that an edge has color $\mathbf{u}\in \mathcal{C}$ if it connects two distinct faces of color $\mathbf{u}$. We will also say that an edge has color $\mathbf{u}$ if it connects a face of color $\mathbf{u}$ and the $\mathbf{u}$-colored boundary, where we define our convention for coloring the boundaries below.

	The lattice we are interested in, shown in Fig.~\ref{Fig:ColorCode} is embedded on a triangle. The three sides of the triangle support three distinct boundaries that are also specified by colors of set $\mathcal{C}$ where the qubits of the boundary of color $\mathbf{u} \in \mathcal{C}$ touch no faces of color $\mathbf{u}$. In Fig.~\ref{Fig:ColorCode} we outline the boundaries with their respective color. Let us also assign a color to each of the three corners of the lattice. We say that a corner of the lattice is colored $\mathbf{u}$ if its vertex supports only one face of color $\mathbf{u}$. We find the $\mathbf{u}$-colored corner at the point where the two boundaries of color $\mathbf{v}$ and $\mathbf{w}$ overlap such that $\mathbf{u} \not= \mathbf{v} \not=\mathbf{w} \not= \mathbf{u}$.

	The color code is such that a qubit is placed on each vertex $v$ of the three-colorable lattice. Quantum error-correcting codes are designed to protect a subspace of the Hilbert space of the total system from common errors. We call this subspace the code subspace, or just the codespace for short. We specify the codespace using the stabilizer formalism. The stabilizer group $\mathcal{S}$ is an Abelian subgroup of the Pauli group acting on $n$ qubits. The code subspace is the $+1$ eigenvalue eigenspace of all of its elements, i.e., $s |\psi \rangle = |\psi\rangle$ for all $s \in \mathcal{S}$ where the code subspace is spanned by state vectors $|\psi\rangle$. The stabilizer generators of the color code are associated to the faces of the lattice. Each face supports two stabilizers $S^X_f = \prod_{v \in \partial f} X_v $ and $S^Z_f = \prod_{v \in \partial f} Z_v$ for all $f$ where $\partial f$ is set of qubits that lie on the boundary of face $f$, and $X_v$ and $Z_v$ are the standard Pauli matrices that act on the qubit on vertex $v$. We will only be interested in the Pauli-Z stabilizers, $S_f^Z$, in this work. As such we will omit the superscripts used for the complete definition and write the relevant stabilizers  more simply as
	\begin{equation}
		S_f = \prod_{v \in \partial f} Z_v,
	\end{equation}
	where again $\partial f$ is the the set of qubits that touch $f$. We show an example stabilizer $S_f = S_f^Z$ in Fig.~\ref{Fig:ColorCode}(a).

	The color code as defined above encodes a single-qubit with an odd code distance $d$ using $n = 3(d-1)(d+1)/4 + 1$ physical qubits. Its low-weight logical operators have string-like support that terminate at each of the three distinct boundaries of the code. Moreover, these string-like logical operators may also branch, see an example of a logical operator at Fig.~\ref{Fig:ColorCode}(b). It will also be convenient to define the logical operators
	\begin{equation}
		\overline{X}_\mathbf{u} =\prod_{v \in \delta \mathbf{u}} X_v ,\quad \overline{Z}_\mathbf{u} = \prod_{v \in \delta \mathbf{u}} Z_v, \label{Eqn:Logical}
	\end{equation}
	where we take the product over the $d$ vertices $\delta \mathbf{u}$ that lie on the boundary of color $\mathbf{u}$. Note that each of the logical operators $\overline{Z}_\mathbf{r}$, $\overline{Z}_\mathbf{g}$ and $\overline{Z}_\mathbf{b}$ are equivalent up to multiplication by an element of the stabilizer group, and so too are the logical Pauli-X operators $\overline{X}_\mathbf{u}$. Of course, $\overline{X}_\mathbf{u} \overline{Z}_\mathbf{v} = - \overline{Z}_\mathbf{v} \overline{X}_\mathbf{u}$ for all colors $\mathbf{u}$ and $\mathbf{v}$.

	A quantum error-correcting code is designed to protect the state encoded in the code space. Let us briefly look at how the color code responds to errors. We will focus on bit-flip errors throughout this work. We write Pauli errors $E = \prod_{v \in E} X_v$ where, by abuse of notation, $E$ denotes both the subset of vertices that support error $E$, as well as the Pauli operator $E$ itself. We measure the stabilizer generators to obtain information about $E$ to find a correction $C$ such that $CE \in \mathcal{S}$. By the definition of the stabilizer group, this correction will recover the encoded state $|\psi\rangle$ that has suffered error $E$, i.e., the state $E |\psi\rangle$ that does not necessarily lie in the code subspace. 
	We say that there is a defect at $f$ if $S_f E = (-1)E S_f $, and the error syndrome is the list of faces that support a defect for error $E$. We also assign each defect a color from $\mathcal{C}$ according to the color of the face on which it lies. 
	A decoding algorithm is designed to determine a correction operator $C$ such that $CE \in \mathcal{S}$ with high probability by taking the syndrome data and prior information about the error model as input. 
	
	Let us finally look at the syndrome produced by some small errors acting on the color code. In Fig.~\ref{Fig:ColorCode}(c) we show a single bit-flip error that creates a single defect on each of its three neighbouring faces. The structure of the code is such that the three defects are all differently colored. Errors can be combined to create longer strings with defects at their endpoints. Fig.~\ref{Fig:ColorCode}(d) shows a string of two errors that lie on the two vertices of a blue edge. This error has created two blue defects at either end of the string. In general, we can say that a string-like error has color $\mathbf{u}$ if it is supported on a sequence of $\mathbf{u}$-colored edges.

	The error syndrome appears differently at the boundary of the lattice. Strings of color $\mathbf{u} \in \mathcal{C}$ can terminate at the $\mathbf{u}$- boundary or the $\mathbf{u}$-colored corner without producing a defect. We show a blue defect terminating at the blue boundary in Fig.~\ref{Fig:ColorCode}(e). Errors can compound further to create defects that are separated over a longer distance. Fig.~\ref{Fig:ColorCode}(f) shows a string-like error where a red, green and blue string all meet at a branching point. The error has created a red and a green defect at the endpoints of their respectively colored string. The blue defect has terminated at the blue boundary.

	\subsection{Symmetries and decoding}
	\label{SubSec:Symmetries}
	
	Here we discuss the symmetries of the color code. The symmetries of the code give us a natural way for a decoder to interpret the syndrome data.

	We define a symmetry~\cite{Brown20parallelized, Tuckett20} as a subset of stabilizers $ \Sigma $ whereby 
	\begin{equation}
		\prod_{s \in \Sigma} s = \openone. \label{Eqn:Symmetry}
	\end{equation}
	This definition of a symmetry reveals a structure among the defects of the error syndrome that allows us to employ minimum-weight perfect matching for decoding. To see why, let us write the eigenvalue of $s$ as $\sigma_s = \pm 1$. Given that $ \left( \prod_{\Sigma} s \right) |\psi \rangle = |\psi \rangle $ due to Eqn.~(\ref{Eqn:Symmetry}), it follows that
	\begin{equation}
		\prod_{s \in \Sigma} \sigma_s =  1.
	\end{equation}
	A direct consequence of this relationship is that there must be an even number of defects, that is, stabilizers with $\sigma_s=-1$, detected among the stabilizers $s \in \Sigma$. More explicitly, this means that every error will give rise to an even number of defects if we restrict our attention to the stabilizers of a symmetry. We can therefore predict the locations of errors by pairing defects that are likely caused by errors drawn from the given error model.  We can regard this as a defect parity conservation law of the error syndrome. This observation is particularly intuitive in topological codes~\cite{Kitaev03, Dennis02, Bombin06, Bombin07a, Brown20parallelized, Nixon21} where we have a local structure among the stabilizer operators. In such cases, errors can be interpreted as strings where defects appear at their endpoints.

\begin{figure}
		\includegraphics{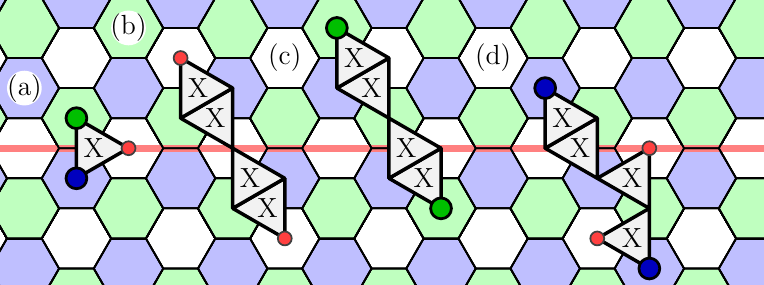}

		\caption{\label{Fig:ColorCodeSymmetry} Decomposing a symmetry of the color code. We show four errors~(a-d). We color the faces of the lattice that are members of $\Sigma_\mathbf{r} = \Sigma_\mathbf{r}^A \cup \Sigma_\mathbf{r}^B$ where we separate the faces of the symmetry into disjoint subsets where $\Sigma_\mathbf{r}^A$($\Sigma_\mathbf{r}^B$) are the faces of $\Sigma_\mathbf{r}$ that lie above(below) the red line. Red faces are excluded from the symmetry. The red line that extends from the left to the right of the figure shows the qubits that support a logical operator of the color code. Errors (a),~(c) and~(d) each give rise to a single defect on either side of the red line on either blue or green faces. Error~(b) creates only red defects, and as such they do not appear on the faces of the symmetry.}
	\end{figure}

	Let us now look at the symmetries of the color code. For now we will consider either infinite or periodic boundary conditions for simplicity. Focusing on just the bit-flip noise model, the symmetries of interest consist of stabilizer operators associated to faces of two specific colors. Let us define the red, green and blue symmetries, $\Sigma_\mathbf{r}$, $\Sigma_\mathbf{g}$ and $\Sigma_\mathbf{b}$, where
	\begin{equation}
		\Sigma_\mathbf{u} = \{ S_f \in \Sigma_{\mathbf{u}} : S_f \in \mathcal{S},\quad  \mathbf{col}(f) \not= \mathbf{u} \}.
	\end{equation}
	For instance, $\Sigma_\mathbf{r}$ contains stabilizer operators $S_f$ for all green and blue faces, see Fig.~\ref{Fig:ColorCodeSymmetry}. Indeed, it is readily checked that the product of all of the colored hexagons on the lattice shown in the figure multiply to give $\openone$.

	In addition to the stabilizers that are members of $\Sigma_\mathbf{r}$, Fig.~\ref{Fig:ColorCodeSymmetry}(a-d) shows four errors that, of course, must respect the symmetry of the code. As such they all give rise to an even number of defects on the faces of $\Sigma_\mathbf{r}$. Fig.~\ref{Fig:ColorCodeSymmetry} also shows the support of a logical operator $\overline{Z}$ by the red line that runs from left to the right along red edges of the lattice.

	We now consider how we can use symmetries to find a correction operator. The problem of decoding can be reduced to estimating the commutator of $E$ and the logical operators of the code. Recall that we seek a correction operator $C$ such that $CE \in \mathcal{S}$. We begin by propsing a trivial correction operator $C'$ that restores the code to any state in the code space. Such an operator is easy to evaluate for topological codes by, say, finding a collection of string-like operators that move all of the defects to some common point on the lattice. We can then ask if $C'$ has the same commutator as $E$ with respect to the logical operator $\overline{Z}$. If we estimate that their commutators are the same, then we can choose $C = C'$ to recover the encoded state. Otherwise, we choose $C = \overline{X} C'$ to recover the encoded state. It therefore remains to determine the commutator of $E$ with $\overline{Z}$.

	Using the setup presented in Fig.~\ref{Fig:ColorCodeSymmetry} we see that errors that anti commute with $\overline{Z}$ produce a single defect on either side of the logical operator. In fact, we find that errors make an odd parity of defects on either side of the logical operator if and only if the error anti commutes with the logical operator. We make this claim rigorous in Appendix~\ref{App:SymmetriesAndLogicals}. We therefore find that we can determine the commutator of $E$ with $\overline{Z}$ by pairing nearby defects over the entire lattice and then counting the number of pairs of defects that are matched across the red line that supports the logical operator. The number of pairs that straddle this line will give us the parity of the number of qubits that the logical operator shares with $E$. This number gives us the commutator between $E$ and $\overline{Z}$ whereby an even(odd) parity of edges crossing the support of the logical operator imply that $E$ commutes(anti commutes) with $\overline{Z}$, thus allowing us to evaluate $C$ and complete the decoding problem. In what follows it remains to explain how we use matching to determine a likely error that produced the syndrome to determine $C$.

	\subsection{A minimum-weight perfect-matching decoder and the restricted lattice}

	\label{SubSec:MWPM}

	We can use minimum-weight perfect matching to find the commutator between some logical operator and an error that was likely to have caused the syndrome. As we have explained in the previous subsection this is sufficient to find a correction operator. The minimum-weight perfect-matching algorithm~\cite{Edmonds65, Kolmogorov09} takes as input a graph with weighted edges and returns a perfect matching, i.e., a graph where all the vertices of the input graph are connected by exactly one edge of the input graph, such that the sum of the weights of the edges of the matching are minimal. Its complexity is $O(V^3)$ where $V$ is the number of vertices for the input graph. On the two-dimensional lattice we expect $V = O(p d^2)$ giving a worst case runtime something like $O(d^6)$. See Ref.~\cite{Dennis02} where this idea was first employed for decoding topological codes. Let us also remark on recent work detailing a Python implementation of the algorithm~\cite{Higgott21}. Here we explain how we use minimum-weight perfect matching to decode the color code using the symmetries we have illustrated in the previous subsection.

	\begin{figure}
		\includegraphics{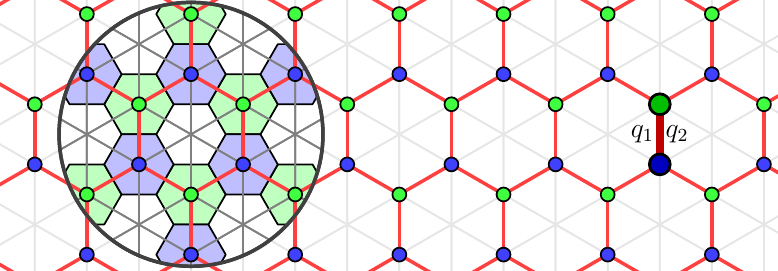}
		
		\caption{The $\mathbf{gb}$ restricted lattice from the symmetries of the color code. We draw the color code on the dual lattice where faces are replaced by vertices and qubits are replaced by triangles. We highlight the stabilizer vertices associated to the symmetry with blue and green vertices. We overlay the dual lattice with the primal lattice in the large circle to the left of the figure. Two adjacent stabilizers are connected by an edge. Each edge has two adjacent qubits that can be flipped to create defects at the endpoints of each edge, see e.g., $q_1$ and $q_2$ to the right of the figure. If exactly one of these two qubits are flipped then two defects on the bold vertices will be created on the restricted lattice. \label{Fig:RestrictedLattice}}
	\end{figure}

	We perform minimum-weight perfect matching to pair defects on a restricted lattice; a concept first introduced in Ref.~\cite{Delfosse14}. Let us see now how the symmetries of the color code give rise to the restricted lattices proposed in Ref.~\cite{Delfosse14}, see Fig.~\ref{Fig:RestrictedLattice}. The figure shows the color code model on the dual lattice, where qubits lie on the triangles of the lattice, and the stabilizers are represented by vertices. The restricted lattice is obtained by projecting each of the qubits supported on a triangle onto an edge that connects, say, the blue and the green vertices that are adjacent to the respective qubit. The restricted lattice is shown by red edges in Fig.~\ref{Fig:RestrictedLattice}. Let us now see how the symmetry relates to the restricted lattice. To the left of Fig.~\ref{Fig:RestrictedLattice} we show the dual lattice overlaid with the primal lattice that we have already defined.  The figure highlights the blue and green vertex stabilizers that correspond to the stabilizers of the  $\Sigma_\mathbf{r}$ symmetry. Let us then say that the restricted lattice corresponding to this symmetry is the $\mathbf{gb}$-restricted lattice. In general we say that the restricted lattice that corresponds to the symmetry $ \Sigma_\mathbf{u}$ is the $\mathbf{vw}$-restricted lattice with colors $\mathbf{u}\not= \mathbf{v} \not=\mathbf{w} \not=\mathbf{u}$.

	Let us now consider how errors appear on the restricted lattice. Every qubit is adjacent to exactly two highlighted vertices; one green and one blue. Therefore, a single qubit error will produce two defects separated by a single edge.  In general, the smallest number of errors that are required create a pair of defects on this symmetry that are separated by a distance $w$ along the edges of this restricted lattice must have weight of at least $w$. It is also worth remarking that there are two qubits projected onto each edge, for instance, both qubits $q_1$ and $q_2$ are projected on the highlighted thick red edge. An error on either $q_1$ or $q_2$ will create a pair of defects that lie at the endpoints of the highlighted edge on the restricted lattice. The restricted lattice we have obtained was likened~\cite{Delfosse14a} to the surface code where qubits lie on the edges of the lattice~\cite{Kitaev03}. The surface code on the hexagonal lattice has been considered explicitly in Refs.~\cite{Fujii12a, Delfosse14a, Chubb21}.

	We can estimate a least-weight correction with respect to a symmetry using minimum-weight perfect-matching. We produce a graph where we assign each defect on the restricted lattice a vertex. We then produce a complete graph where edges are assigned a weight that is proportional to the separation of the defects along the shortest path over the restricted lattice. The output of the minimum-weight perfect-matching algorithm indicates a set of low-weight error strings that are likely to have created the error syndrome with respect to the symmetry. Counting the number of edges that pair defects over the support of the logical operator gives us an estimate of the parity of errors supported on the logical operator of interest. One can prove that the probability of estimating the support of the error on the logical operator incorrectly will decay exponentially quickly with $d$ for a sufficiently low error rate using similar arguments to those presented in~\cite{Dennis02}. In Appendix~\ref{App:MWPM} we justify why the solution to the minimum-weight perfect matching algorithm will propose a likely error, and we give details on how we can evaluate the separation between two points on the hexagonal restricted lattice in Appendix~\ref{App:MeasuringEdgeWeights}.

	\subsection{Prior work}
	
	We conclude this preliminary Section by discussing earlier work that has used variations of the decoding strategy above~\cite{Wang10,  Delfosse14, Stephens14, Delfosse17, Aloshious18, Li18, Kubica19, Chamberland20, Beverland21}. In~\cite{Delfosse14} it was shown that the edges obtained from matching on the three restricted lattices we have defined above can be used to find the border of a correction that is consistent with the error syndrome. This decoder produced a threshold of $\sim 8.7\%$ on the hexagonal lattice with periodic boundary conditions which is consistent with the earlier work in Ref.~\cite{Wang10} where a decoder was proposed by consideration of the fusion rules of the anyon model of the color code~\cite{Bombin06}. This number is also aligned with the work of~\cite{Fujii12a} where a threshold of $\sim 15.9\%$ is obtained by matching Pauli-Z errors on the surface code on a hexagonal lattice. Indeed, equivalent values are obtained if we equate this threshold with $2 p (1-p) = 0.159 $ given that, in the color code picture, there are two distinct qubits that can cause an error that will create a pair of defects. 
	
	The matching decoder on the restricted lattice was simplified in~\cite{Kubica19} where it was shown that a local correction can be found using the result of the matching on two of the three restricted lattices. Ref.~\cite{Kubica19} obtained a threshold of $\sim 10\%$ by focusing on two specific restricted lattices on an alternative color code lattice. In a sense, the alternative perspective we have provided here offers another simplification, where we can decode individual logical operators separately by concentrating on the matching found from a single restricted lattice. Generalisations of this decoder have been obtained for the color code undergoing circuit noise that occurs as stabilizer readout is performed~\cite{Stephens14, Li18, Chamberland20, Beverland21} by extending the error syndrome in the temporal direction~\cite{Dennis02, Brown20parallelized}. These examples also generalise the restricted lattice by considering the case where the color code has boundaries. See also~\cite{Li18} where the problem of decoding the color code undergoing bit-flip noise is likened to decoding errors on a Majorana code. 
	
Let us also comment on thresholds obtained with maximum-likelihood decoding. Using a statistical-mechanical mapping that determines the performance of a maximum-likelihood decoder a threshold for bit-flip noise~\cite{Katzgraber09, Katzgraber10} has been obtained as $~\sim 10.9\%$. Thresholds of $\sim 18.9\% $ have been obtained for the color code undergoing depolarising noise~\cite{Bombin12}. These results have been reproduced with a tensor-network decoder~\cite{Bravyi14, Tuckett19} that approximates maximum-likelihood decoding. Fault-tolerant thresholds for a phenomenological noise model of $\sim 4.8\%$ have also been obtained using statistical mechanical modelling~\cite{Andrist11, Katzgraber13, Andrist16}. These remarkably high numbers motivate the development of efficient fault-tolerant decoders for the color code. We summarise threshold results for different color code lattices undergoing a bit-flip noise model in Table~\ref{Tab:Thresholds}.

\begin{table}
\begin{tabular}{|l|l|r|}
	\hline Lattice & Decoder & Threshold \\
	\hline Hexagon & Optimal & $10.9 \%$ \cite{Katzgraber09}  \\
	& Tensor Network & $10.9 \%$ \cite{Chubb21} \\
	& Neural Network & $ 10.0 \%$ \cite{Maskara19} \\
	& \textbf{M\"obius MWPM} & \textbf{9.0\%} \\
	& Restriction MWPM & $8.7 \%$ \cite{Delfosse14} \\
	& Union-Find & $8.4 \%$ \cite{Delfosse17} \\
	& Renormalization-Group & $7.8 \%$ \cite{Sarvepalli12} \\
	\hline Square-Octagon & Optimal & $10.9 \%$ \cite{Katzgraber10} \\
	& Least-weight correction & $10.6 \%$  \cite{Landahl11}\\
	& Restriction MWPM & $10.2 \%$ \cite{Kubica19} \\
	& Union-Find & $9.8 \%$ \cite{Kubica19}  \\
		& Renormalisation-Group & $8.7 \%$ \cite{Bombin12}  \\
	\hline
\end{tabular}
\caption{\label{Tab:Thresholds} Thresholds obtained for different color codes undergoing bit-flip noise. The result obtained in the present paper using minimum-weight perfect matching(MWPM) on a unified M\"obius lattice is shown in boldface. This decoder is described in Sec.~\ref{Sec:Decoder}.}
\end{table}
	
	\section{Decoding on the M\"obius strip}
	\label{Sec:Decoder}
	
	Let us now describe our decoder. We find that we can decode the color code using a single minimum-weight perfect-matching subroutine on a lattice that is embedded on a M\"obius strip. In what follows we will examine the symmetries of the color code to show how we arrive at the decoder we present. We will go on to explain how the decoder overcomes the challenges of decoding the color code on a triangular lattice. We also explain how our decoder deals with the issue of degeneracy that arises when looking at the syndrome on a restricted lattice.

	\subsection{Symmetries of the color code with boundaries}

\begin{figure}
		\includegraphics[width=\columnwidth]{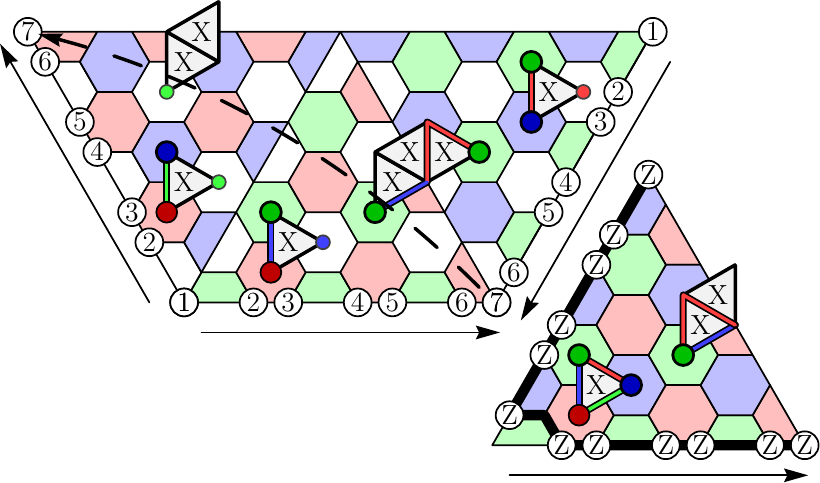}
		\caption{\label{Fig:Mobius} The M\"obius symmetry. An image of the error syndrome is shown three times on three different restricted lattices. The restricted lattices are connected at their boundaries to reconstruct a single unified lattice that is embedded on a M\"obius strip. We number unified qubits to the left and right of the lattice. Note that the qubits are oppositely aligned on each side of the lattice, thus giving the restricted lattice M\"obius topology. We show the image of the error syndrome shown to the right of the figure on the M\"obius symmetry. We also show the operator $b_{\mathbf{g}}$, which is the product of all the stabilizers on the red and blue faces.}
	\end{figure}

	Here we look more closely at the color code symmetries at the boundaries of the lattice to show the construction of the single M\"obius symmetry. In SubSec.~\ref{SubSec:Symmetries} we found the restricted lattices used for a minimum-weight perfect-matching decoder by identifying that the product of all the faces on two of the three colors of the lattice give rise to a symmetry. However, this is not true on the lattice with boundaries. We define a boundary operator $b_\mathbf{u}\in \mathcal{S}$ such that
		\begin{equation}
		b_\mathbf{u} = \prod_{ \mathbf{col}(f) \not= \mathbf{u}} S_f,
	\end{equation}
	where we take the product of all the faces that are not colored $\mathbf{u}$. As an example we show $b_\mathbf{g}$ to the right hand side of Fig.~\ref{Fig:Mobius}. We note also that $b_\mathbf{u}  = \overline{Z}_\mathbf{v} \overline{Z}_\mathbf{w} $, where $\mathbf{u} \not= \mathbf{v} \not= \mathbf{w} \not= \mathbf{u}$ where we defined these instantiations of the logical operators in Eqn.~(\ref{Eqn:Logical}).
	
	Before explaining the construction of the M\"obius strip, it will be helpful to first show how we can recover the standard restricted lattices of the color code with boundaries.
	Indeed, the inclusion of the operator $b_\mathbf{u}$ in the symmetry $\Sigma_\mathbf{u}$ together with the stabilizers $S_f$ on faces with color $ \mathbf{col}(f) \not= \mathbf{u}$ give us a symmetry that enables us to decode with matching. Specifically, we have that $\prod_{s \in \Sigma_{\mathbf{u}}} s = \openone$ if we take symmetries
	\begin{equation}
	\Sigma_\mathbf{u} = \left\{ b_\mathbf{u} , \left\{ S_f \right\}_{\mathbf{col}(f) \not= \mathbf{u}}  \right\},
	\end{equation}
	for the color code with boundaries.
	In practice, the addition of this operator means that we can match defects on the faces $\Sigma_\mathbf{u}$ onto one of the two boundaries with color not equal to $\mathbf{u}$. This strategy is commonly adopted elsewhere in the literature, see for instance~\cite{Stephens14, Chamberland20, Chamberland20a, Beverland21}. Explicitly considering the boundary operators reveals additional structure between these restricted lattices. We find that
	\begin{equation}
		b_\mathbf{r} b_\mathbf{g} b_\mathbf{b} = \openone. \label{Eqn:BoundaryRelation}
	\end{equation}

Let us look at these correlations from a physical perspective. This will motivate the method of decoding we propose. As we have already discussed, the symmetries of the color code correspond to a $\mathbb{Z}_2 \times \mathbb{Z}_2$ conservation law among the defects of the color code in the bulk. To see this, one can check that there is no single qubit error acting on the bulk of the lattice that will violate the relation
	\begin{equation}
		|\#\mathbf{r}|_2 = |\#\mathbf{g}|_2 = |\#\mathbf{b}|_2, \label{Eqn:ConservationLaw}
	\end{equation}
	where $|\#\mathbf{u} |_2$ denotes the number of defects of color $\mathbf{u}$ modulo 2. Equivalently, we can write the conservation law in terms of stabilizer operators such that
	\begin{equation}
		b_\mathbf{r} |\phi\rangle =  b_\mathbf{g}   |\phi\rangle =  b_\mathbf{b}   |\phi\rangle.
	\end{equation}
	for states $|\phi\rangle = E_\text{bulk} |\psi \rangle$ where Pauli errors $E_\text{bulk}$ act on qubits on the bulk of the lattice of code states of the color code $|\psi\rangle$. This is clear because the boundary operators $b_\mathbf{u}$ are not supported in the bulk of the lattice.

	Errors on the boundary of the lattice effectively violate the global defect conservation law shown in Eqn.~(\ref{Eqn:ConservationLaw}). The boundary operators record these violations. Let us take, for example, the error shown in Fig.~\ref{Fig:Mobius}. A single error on the green boundary means that $|\#\mathbf{r}|_2 = |\#\mathbf{b}|_2$ but  $|\#\mathbf{r}|_2  \not= |\#\mathbf{g}|_2 $ and $|\#\mathbf{b}|_2  \not= |\#\mathbf{g}|_2 $. Likewise, and correspondingly, we have that $b_\mathbf{r} |\xi\rangle =  b_\mathbf{b} |\xi\rangle $, but that $b_\mathbf{r} |\xi\rangle \not=  b_\mathbf{g} |\xi \rangle $ and  $  b_\mathbf{b} |\xi\rangle \not=  b_\mathbf{g}   |\xi\rangle $, where now $| \xi \rangle = E_\text{bdry.} |\psi \rangle$ for the boundary error $E_\text{bdry.}$ shown in Fig.~\ref{Fig:Mobius}.

	We have thus seen that the defect parity conservation on the $\mathbf{gb}$- and $\mathbf{rg}$-restricted lattices are violated if and only if a green defect is created from the green boundary or from the green corner. Similarly, a red(blue) defect created at the red(blue) boundary or corner qubit will simultaneously violate the defect parity conservation law on the $\mathbf{rb}$- and $\mathbf{rg}$- ($\mathbf{gb}$- and $\mathbf{rb}$-)restricted lattices. 
	
	It is important to find a correction that respects global defect conservation at the boundaries of the color code However, a problem we find by considering the defect configuration on restricted lattices independently is that we can obtain corrections that do not respect Eqn.~(\ref{Eqn:BoundaryRelation}).

	We find that we can obtain a correction that respects the boundary operators by combining the restricted lattices to produce a single unified lattice. We show the construction of the new lattice in Fig.~\ref{Fig:Mobius}. We show all three restricted lattices combined along their boundaries. We call this join between two boundaries of the restricted lattice a crease, and we give each crease a color, $\mathbf{u} = \mathbf{r},\, \mathbf{g},\, \mathbf{b}$, according to the color of the logical operator $\overline{Z}_\mathbf{u}$ its qubits support; see Eqn.~(\ref{Eqn:Logical}). For instance, we see that the central $\mathbf{rg}$-restricted lattice is combined with the $\mathbf{gb}$-restricted lattice at the right of the figure along the green crease. Note also that the $\mathbf{gb}$-restricted lattice is a reflection of the $\mathbf{rg}$-restricted lattice over the green crease. The qubits at the green corner of these two lattices are unified. 

In the same way, we unify the blue boundary and the blue corner of the $\mathbf{rb}$- and $\mathbf{gb}$-restricted lattices, and we unify the red boundary and the red corner of the $\mathbf{rg}$- and $\mathbf{rb}$-restricted lattices to obtain the single restricted lattice shown in the figure. The resulting lattice gives rise to a symmetry that respects defect conservation symmetry among its boundary terms. We call this lattice the unified lattice, as it combines all three restricted lattices. For the triangular lattice we have introduced we find that our unified lattice is supported on a M\"obius strip. This unified lattice is an interesting example where its corresponding symmetry includes all of the stabilizer generators $S_f$ twice. A similar idea was used in Ref.~\cite{Tuckett20} to find a decoder for the tailored surface code undergoing biased errors.

	\subsection{Assigning weights to the edges of the matching graph on the unified lattice}
	\label{SubSec:EdgeWeights}

	\begin{figure}
		\includegraphics{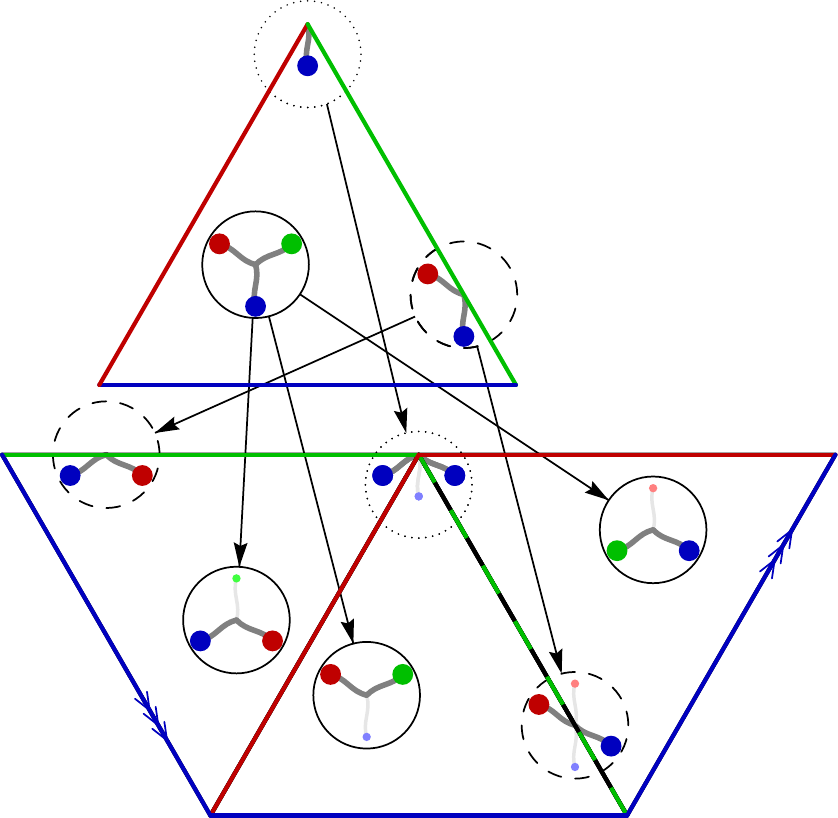}
		\caption{\label{Fig:Edges} The error syndrome mapped onto the M\"obius strip. At the top of the figure we show three single qubit errors; one in the bulk that gives rise to three defects, one at the boundary that creates two defects, and an error at the corner where only one defect is produced. They are circled with solid, dashed and dotted lines respectively. Arrows point to the image of the error and its syndrome on the M\"obius strip. We observe that the single-qubit error in the bulk produces three edges on the M\"obius strip. A single-qubit error on the boundary produces two edges; one in the bulk of the lattice, and one that crosses the crease that represents the logical operator supported on the green boundary. An error at the corner of the color code produces a single edge on the M\"obius strip.}
	\end{figure}

	Here we explain how to decode the color code using the new symmetry. 
	We look at how errors, and their syndrome, map onto the unified lattice. We will look at different single qubit errors to explain how we assign weights to the edges of the input graph, and we will explain how we determine the support of the error on a logical operator using the intuition we presented in SubSecs.~\ref{SubSec:Symmetries} and~\ref{SubSec:MWPM}.

	Errors map differently onto the unified lattice depending on whether the error occurred in the bulk, on the boundary, or at a corner of the lattice, see Fig.~\ref{Fig:Edges}. We propose that a good strategy to this end is to assign weights to unit edges, i.e. edges created by single qubit errors, such that the sums of the weights of all the edges associated to each single qubit error are equal. Let us begin by looking at a single-qubit error in the bulk of the lattice. As we see in the solid circles in Fig.~\ref{Fig:Edges} this error produces three separate edges on the unified lattice. Without loss of generality then, let us assign a weight of $1$ to a single edge in the bulk of the unified lattice. The sum of all the edges that identify the single qubit error then is $3$.

We next consider an error at the boundary of the lattice, such as that circled by the dashed line in Fig.~\ref{Fig:Edges}. We see that this error produces two edges on the unified lattice. These two edges are distinct in the following sense: the edge to the left of the M\"obius strip can also be created by an error in the bulk. On the other hand, the edge to the right that crosses the crease is not consistent with any bulk error. We have already assigned a weight of 1 to all edges created by a single-qubit error in the bulk. We therefore give edges that cross the crease a weight of $2$. This choice is such that the sum of the weights of the edges associated to the error at the boundary is equal to the sum of the weights of the edges associated to the error in the bulk: $3$.

	We finally observe that the error produced at the corner creates a single edge; see the error circled by the dotted line in Fig.~\ref{Fig:Edges}. We must assign a weight of $3$ to this edge that passes through the corner then to ensure consistency with the other single qubit errors. As errors compound to make longer strings, we assign weights to the edges that pair two well-separated defects as to the sum of the weight of the unit edges along the shortest path connecting the two defects, where the unit edges along this path take weights according to the assignment we have proposed above.

We use the resulting matching to find the commutator with some specified logical operator. Following the arguments we have given in the previous section, the parity of the number of edges that cross the dashed line on the green crease is consistent with the commutator of the errors with the logical operator supported on the green boundary: $\overline{Z}_\mathbf{g}$. One can easily check that all the single-qubit errors that lie on the green boundary, including the red and blue corners, each create a single edge on the M\"obius strip that crosses this line, whereas no other single-qubit errors produce any edges that cross this line. We therefore count all the edges that cross this line and thereafter propose a correction operator that is consistent with the commutator that is learned from the matching.

	\subsection{Correcting high-weight errors}
	\label{SubSec:HighWeightErrors}

	\begin{figure}
		\includegraphics{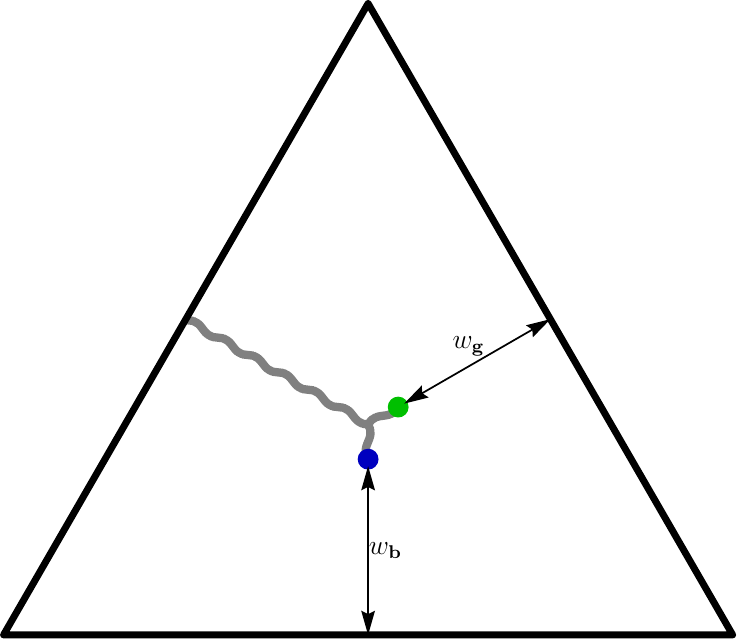}
		\caption{\label{Fig:dOnThreeError} A difficult error for a decoder that does not account for correlations between green and blue defects. For a sufficiently large code, with $d \gtrsim 11$, one can find errors of weight $ w  = O(d/3) $ such that $w  \ll (d-1) / 2$ where, if we neglect the position of the blue defect, the green defect is paired to the green boundary with a correction of weight $w_\mathbf{g} < w$. Likewise, a decoder that matches on the $\mathbf{rb}$ lattice will suggest a correction where the blue defect is paired to the blue boundary with an operator of weight $w_\mathbf{b} < w $. These choices will lead the decoder to fail.}
	\end{figure}

	In this SubSection and that which follows we motivate our choice to decode using the symmetry on the unified lattice. To do so, we compare our decoder to a na\"ive implementation of a decoder that finds a correction using a matching on two restricted lattices to identify the challenges that arise when designing a decoder for a color code with boundaries. Without loss of generality we assume the decoder finds a correction using the $\mathbf{rb}$- and $\mathbf{rg}$-restricted lattices such that the decoder does not identify correlations between green and blue defects.

	We consider the error shown in Fig.~\ref{Fig:dOnThreeError}. It shows an error $E$ of weight $ w = O(d / 3) $ where $w \ll (d-1) / 2$ for lattices with $d \gtrsim 11$ that we might hope to be able to correct. The error extends from the red boundary such that a blue and a green defect are created at the end of the string $E$ near to the centre of the triangle. We also consider operators $C_\mathbf{g}$ and $C_\mathbf{b}$ that pair the green and blue defect to the green of blue boundary, respectively, and we have that $EC_\mathbf{g}C_\mathbf{b}$ is a logical operator. The weights of operators $E$, $C_\mathbf{g}$ and $C_\mathbf{b}$ are $w$, $w_\mathbf{g}$ and $w_\mathbf{b}$, respectively.

	The decoder will match on the $\mathbf{rb}$- and $\mathbf{rg}$-restricted lattices. In the case of the $\mathbf{rg}$-restricted lattice the decoder must decide to pair the green defect onto either the  red boundary or the green boundary. By consideration of the geometry of a triangle, one can find errors of weight $w = d/3 + \textrm{const.}$ with $\textrm{const.}>0$ a small integer such that the matching subroutine will pair the green defect onto the green boundary provided $w > w_\mathbf{g}$. Likewise, if $w > w_\mathbf{b}$, the decoder will incorrectly pair the blue defect onto the blue boundary. We therefore see that low-weight errors with $w = O(d/3)$ can lead to a logical failure if both $w > w_\mathbf{g},\,w_\mathbf{b}$. Nevertheless, this is a bad choice of correction given that it can be that $w \ll w_\mathbf{g} + w_\mathbf{b}$. Ideally, we can find a decoder that can consider all of the defects of the lattice in unison to account for this.

\begin{figure}
		\includegraphics{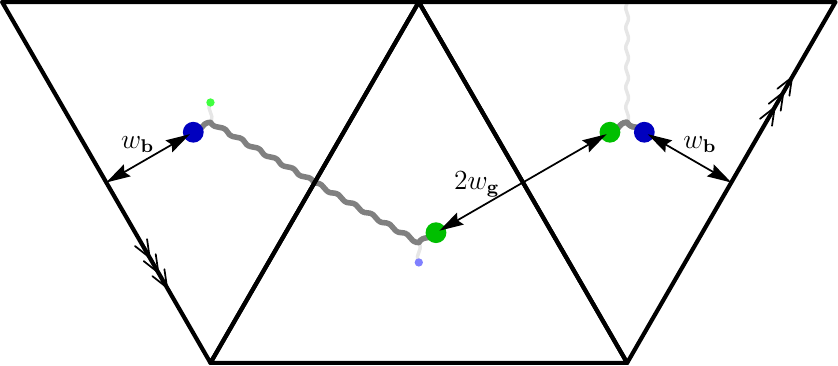}
		\caption{\label{Fig:Unified} The image of the error shown in Fig.~\ref{Fig:dOnThreeError} on the unified lattice. The error, shown in grey, has weight $\sim 2w+1$. However, the distance between the two green defects is $2w_\mathbf{g}$ on the M\"obius strip. Likewise, the distance between the two blue defects is $2w_\mathbf{b}$. We therefore see that the matching decoder will only find the incorrect outcome if $2w+1 \gtrsim 2(w_\mathbf{g} +w_\mathbf{b})$. Given that $w + w_\mathbf{g} + w_\mathbf{b} = d$, we find that the decoder can tolerate errors of the type shown in Fig.~\ref{Fig:dOnThreeError}, i.e., long strings that extend from the boundary, provided $w \lesssim d/2$. }
	\end{figure}

	In Fig.~\ref{Fig:Unified} we consider the same error on the unified lattice. The image of the error effectively doubles its weight to $\sim 2w$. However, to pair the defects incorrectly, the decoder must produce edges of weight $2w_\mathbf{g}$ to connect the green defects that appear on both the $\mathbf{rg}$- and $\mathbf{gb}$-restricted lattices, together with an edge of weight $2w_\mathbf{b}$ to pair the blue defects on the $\mathbf{rb}$- and $\mathbf{gb}$-restricted lattices. As such, the decoder will only fail if
	\begin{equation}
		2w+1 \ge 2(w_\mathbf{g} + w_\mathbf{b}),
	\end{equation}
	where we have added a unit on the left hand side of the inequality to account for the single edge to pair the green and blue defect together on the $\mathbf{gb}$-restricted lattice. As such, we see that matching on the unified lattice enables us to correct high-weight strings that extend from some boundary. Moreover, clearly, the unified lattice is invariant under color exchange, as it accounts for all three restricted lattices equally, as such there is no color dependence on the choice of correction.

	\subsection{Accounting for the degeneracy of errors}
	\label{SubSec:Degeneracy}
\begin{figure}
		\includegraphics{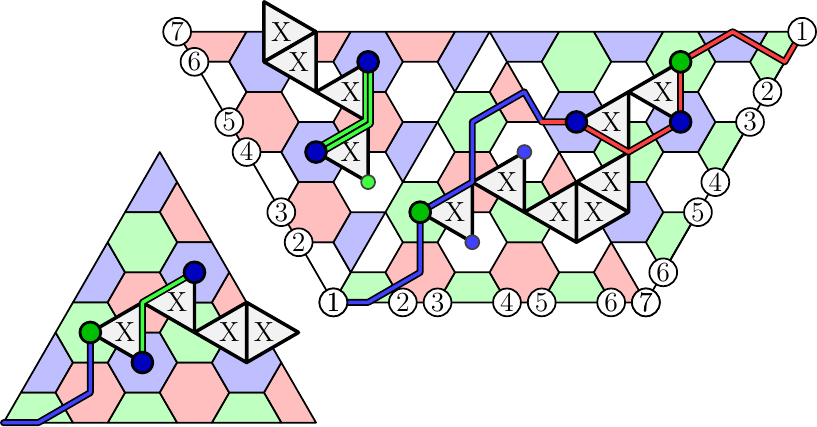}
		\caption{\label{Fig:BadEntropy} Here we consider a weight four error on a $d=7$ color code lattice. The left of the figure shows the error on the color code lattice, together with the correction a na\"ive decoder will choose. Specifically, matching on the $\mathbf{rb}$-restricted lattice will pair the two blue defects correctly. However, matching on the $\mathbf{rg}$-restricted lattice will incorrectly pair the green defect along the shortest path to the green corner of the lattice. On the right we show the image of the same error on the unified lattice. Given that the decoder accounts for information from all three restricted lattices, we find that the decoder will find the right correction by pairing defects along the solid lines. The incorrect matching, shown by the dashed lines, has a higher weight. As such, in this example, a decoder that matches on the unified lattice will be successful.}
	\end{figure}

	Let us now consider issues that arise due to the degeneracy of errors when we consider syndromes on the restricted lattice. As we have already alluded, a problem that can arise when we make a decoder based on matching on a restricted lattice is that some syndrome information is disregarded. Indeed, we can find many different errors that produce different syndromes on the color code, that give rise to the same syndrome on the restricted lattice. Let us consider the error shown in Fig.~\ref{Fig:BadEntropy}. Again, considering the na\"ive decoder, where we find a correction by pairing only on the $\mathbf{rg}$- and $\mathbf{gb}$-restricted lattices, the decoder is likely to pair the green defect to the top corner instead of the green boundary. However, this correction has a weight of $5$, whereas the error itself has a weight of 4. As such, we might expect a better strategy to account for this degeneracy in the syndrome.
	
	In Fig.~\ref{Fig:BadEntropy} we show the image of the error on the unified lattice. Again, as the decoder considers all the syndrome information equally, in this example we find that the decoder will find the correct solution. In the figure we compare the weights of the edges of both the correct and incorrect matching. We find that the correct matching, shown by solid lines, has a weight of 8, whereas the incorrect matching has a weight of 11, where we remember the assignment of weights to edges that pass over boundaries and corners as explained in SubSec.~\ref{SubSec:EdgeWeights}.

	\subsection{Matching with branching errors and finding a low-weight correction}
	\label{SubSec:AltCorrection}

\begin{figure}
		\includegraphics{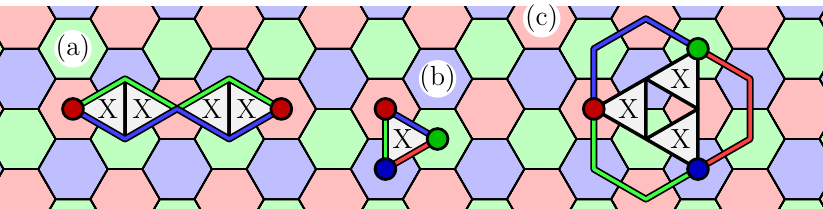}
		
		\bigskip
		
		\includegraphics{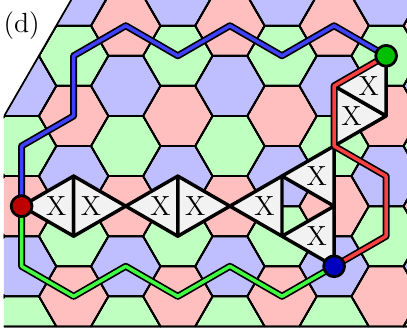}
		\includegraphics{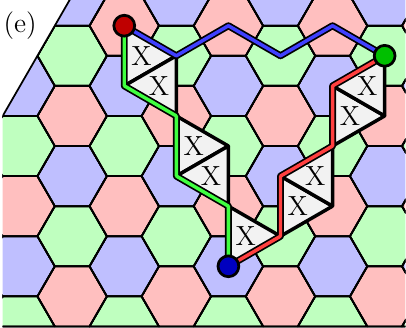}
		
		\caption{\label{Fig:ErrorsAndEdges} Matching found for the syndrome of small errors using the matching algorithm. (a)~Typical string-like errors with weight $w$ are paired by two edges from the matching. The sum of the lengths of the edges that pair these two red defects has length $\ell \sim 2w$. (b)~and (c)~show small errors with weight $w = 1$ and $w=3$, respectively, that are identified by the matching algorithm with three edges in the matching that have total length $\ell = 3w$. (d)~and (e)~show branching errors of weight $w = 9$ whose matchings have lengths $\ell = 2w +3 $ and $ \ell = 3 (w + 1) / 2 $, respectively.}
	\end{figure}

	Unlike the surface code~\cite{Dennis02}, we find that the sum of the lengths of all the edges returned from the minimum-weight perfect matching is not necessarily proportional to the weight of the least-weight correction. In fact we find that the sum of the lengths of edges that indicate a branching error has occurred is quite a complicated function. Without intervention, a minimum-weight perfect-matching decoder will be biased towards a locally minimal solution with weight that is greater than the least-weight correction. Let us look at some errors to illustrate this problem before proposing a solution, see Fig.~\ref{Fig:ErrorsAndEdges}.

	A typical string-like error of weight $w$ will create two defects, one at each of its endpoints, that matched twice on the unified lattice. As such, we have that the total length of the edges from the matching associated to this error will have length $\ell \sim 2 w$, see Fig.~\ref{Fig:ErrorsAndEdges}(a). In contrast, the sum-total of the lengths of the edges that identify a branching error may be larger. For instance Fig.~\ref{Fig:ErrorsAndEdges}(b) shows a single error that is identified with three edges. We can find branching points of three qubits, where the sum total of the weights of the edges that match the defects of the branch is nine, see Fig.~\ref{Fig:ErrorsAndEdges}(c). As such, we observe that the sum of the weights of the edges may have $\ell \sim 3 w$ around a branching error. 
	
	Let us now look at how the sums of the weights of edges that identify an error misalign with the weight of the error. We will consider an error $E_B$ with weight $w_B $ that includes a branch that has three qubits each contributing three edges. The error lies on the support of a least weight logical operator such that there is an alternate correction $E_A$ with weight $w_A = d - w_B $ and $E_A E_B$ is a logical operator.

	We first consider an error with a branch such that the sum total of the lengths of the edges that identify the branch are $\ell_B \sim 2 w_B +3$, see Fig.~\ref{Fig:ErrorsAndEdges}(d). In contrast, its alternate correction $E_A$ is matched with edges of length $\ell_A \sim 2 w_A$. The minimum-weight perfect-matching algorithm will identify $E_A$ as the error if $\ell_A > \ell_B$. With the relations proposed above find that this holds if
	\begin{equation}
		d / 2  - 3 / 4 >  w_B .
	\end{equation}
	We therefore see that there are branching errors with weight $w_B = (d-1) / 2$ that can lead the decoder to choose the alternate correction. However, these errors should be correctable.

	We also find that there are branching points that have very low weight edges. For instance, we find that there are branching errors that are identified by the minimum-weight perfect-matching algorithm with edges of length $\ell_B \sim 3 w_B / 2 $. We show one such error in Fig.~\ref{Fig:ErrorsAndEdges}(e) with $\ell_B \sim 3 w_B / 2 + 3/2$. We find that this can compromise the performance of the decoder significantly. Let us find the largest error $E_A$ such that $ \ell_A < \ell_B$ where again, the weight of $E_A $ is $w_A$, and has edges of length $\ell_A \sim 2w_A$ and $E_A E_B$ is a logical operator of weight $d$. We find that $\ell_B > \ell_A$ with errors $E_A$ of weight
	\begin{equation}
		w_A < 3 d / 7 - 3/7.  \label{Eqn:Branch}
	\end{equation}
	Once again, we find that there are errors with weight $w_A \sim  3d/7 $ where clearly $ w_A \ll (d-1)/2$ for large $d$ that are misidentified due to the low weight of the edges that match the branches similar to those shown in Fig.~\ref{Fig:ErrorsAndEdges}(e).

	Having identified the problem that branching errors have a weight  $ \ell / 3 \lesssim w \lesssim 2 \ell / 3 $ with $\ell$ the sum of the lengths of the edges that identify the branch, we need to perform additional analysis to evaluate the weight of errors that include a branch. The general problem of finding a least-weight correction will require a least-weight hypergraph matching algorithm. 
	
\begin{figure}
		\includegraphics{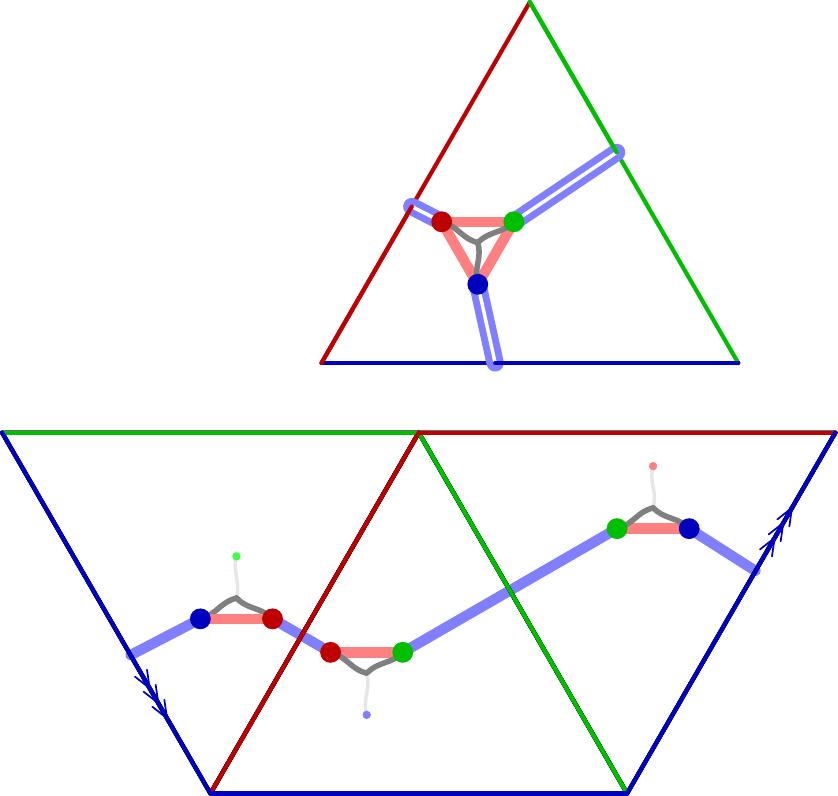}
		\label{Fig:TwoCorrections}
		\caption{Obtaining two low-weight corrections that are logically inequivalent. We perform minimum-weight perfect-matching multiple times. The first time with a single use of the minimum-weight perfect-matching algorithm that we discussed above. We show this matching in blue. In the second use of the subroutine we change the topology of the manifold to force the algorithm to find an alternative low-weight correction, shown in red.}
	\end{figure}

	In lieu of hypergraph matching we propose another solution to find the weight of large branches that may contribute to a bad choice of correction where we use minimum-weight perfect-matching multiple times to find alternative corrections. Our method is based on a similar idea proposed in Ref.~\cite{Hutter14} used to find an alternative correction for the surface code with boundaries. We explain the details of this in Appendix~\ref{App:AlternativeCorrections}. This results in two inequivalent low-weight corrections, see Fig.~\ref{Fig:TwoCorrections}. We can thus use the alternative correction to correct instances where a single use of the minimum-weight matching algorithm is biased towards a higher-weight solution. We discuss how we compare alternative corrections in the following Section; SubSec.~\ref{SubSec:Exhaustive}

\section{Numerical results}
\label{Sec:Results}

	In this Section we simulate the error-correction procedures we have proposed to evaluate their performance. We use an independent and identically distributed noise model to test the low error rate behaviour of the code as well as its threshold. We begin by using a basic implementation of our decoder that we shall henceforth call the M\"obius decoder. This decoder uses the correction that is obtained from a single matching subroutine on the unified lattice. At low error rates, we show that the decoder is able to correct errors with a logical failure rate $\sim p^{\alpha d}$ with $\alpha \approx 3/7$, as predicted in Eqn.~(\ref{Eqn:Branch}), and $p $ is the probability of a bit-flip error occurring on any given qubit. We also evaluate the error tolerance threshold to be $9.0\%$ using the M\"obius decoder. 
	
The value of $\alpha$ we obtain indicates that, as expected, there are errors of weight $\le (d - 1) / 2$ that lead the decoder to fail. As such, we improve the M\"obius decoder by introducing a variant that we term the comparative decoder. This variant uses an additional subroutine to find an alternative correction that is compared with the result obtained by the M\"obius decoder. We verify the comparative decoder's ability to correct errors with weight up to the code distance for $d \le 13$ by conducting an exhaustive search through all possible weight $w \leq (d-1)/2$ error configurations.

	\subsection{Logical failure probability at low error rates}

\begin{figure}
		\includegraphics[width=\columnwidth]{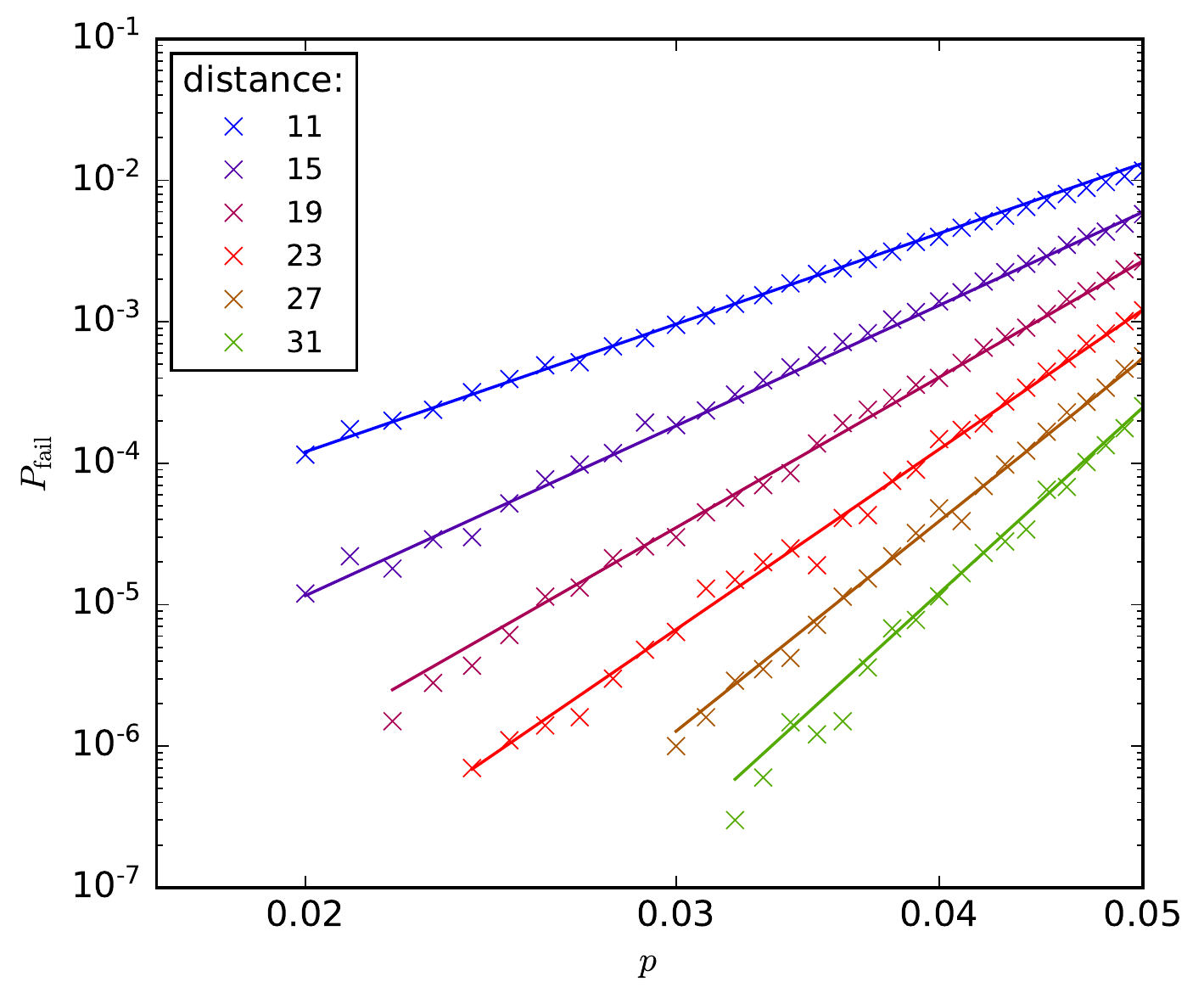}
		\caption{\label{Fig:lowPFit} Plot with logarithmic axes showing the logical failure rate $P_\textrm{fail}$ for the triangular color code as a function of physical error rate $p$ using the M\"obius decoder. We plot the exact form of the ansatz, Eqn.~(\ref{Eqn:ansatz}) for each code distance $d$, showing good agreement with the data. Data points are collected using between $10^6$ and $10^7$ Monte Carlo samples.}
	\end{figure}

	At low physical error rates, we expect the logical failure rate to decay rapidly with code distance. We show our data in Fig.~\ref{Fig:lowPFit}. The major contribution to ${P}_{\text{fail}}$ will be errors of minimal-weight $t \le (d-1) / 2 $ that lead to logical failure. We compare the decoder performance at low ${p}$ values to the ansatz~\cite{Fowler13a}
	\begin{equation}
	{P}_{\text{fail}}=\beta({Np})^{\alpha {d}+\gamma} \label{Eqn:ansatz}
	\end{equation}
	By fitting to the data collected, we get
	$\beta=0.148$, the entropy term ${N}=12.49, \gamma=0.488$, and $\alpha=0.422$. We explain how we obtain these values in Appendix~\ref{App:Fitting}. 	
	Remarkably, the value of $\alpha$ we observe is close to $3/7 (\approx 0.428)$. We predicted the M\"obius decoder may fail for errors of weight $t = 3d / 7$ from our analysis of branching errors in Fig.~\ref{Fig:ErrorsAndEdges}. Specifically, see the argument that gives rise to Eqn.~(\ref{Eqn:Branch}).

	We find that a color code decoder that can correct $t= \alpha d$ errors with $\alpha = 3/7$ is competitive with a decoder for the square-lattice surface code that decodes up to its distance~\cite{Watson12, Bravyi13, Beverland19} using a commensurate number of qubits. Let $n =d_{\textrm{s.c.}}^2$ be the number of qubits on the distance $d_{\textrm{s.c.}}$ surface code. Decoding the surface code up to its distance means we have $t_\textrm{s.c.} = \sqrt{n}/2 $.  We compare this with a color code with an equal number of qubits, $n =  3(d-1)(d+1)/4 + 1 \approx 3d^2 / 4$. Rearranging, we have that $d = 2 \sqrt{n}  /\sqrt{3} $. Substituting this expression into our value of $t = 3 d / 7$ we find
		\begin{equation}
		t = 3d / 7 = 6 \sqrt{n} / 7 \sqrt{3} \approx 0.495 \sqrt{n}, \label{Eqn:dforCC}
	\end{equation}
almost reaching the optimal logical failure rate scaling of the surface code. Together with its capability of performing fault-tolerant logical operations with low overhead, we might expect the color code to require fewer resources for quantum-logic operations at low error rates and modest system sizes. Moreover, given we have not reached the capacity of the color code to correct up to weight $(d-1) / 2$ errors, improved decoders have the potential to outperform the logical failure rate scaling of the surface code. In Sec.~\ref{SubSec:Exhaustive} we show an improved version of our decoder can correct all errors of weight $(d-1)/2$ for $d \le 13$. We first evaluate the threshold of the basic implementation of our decoder.

	\subsection{Thresholds}

\begin{figure}
		\includegraphics[width=\columnwidth]{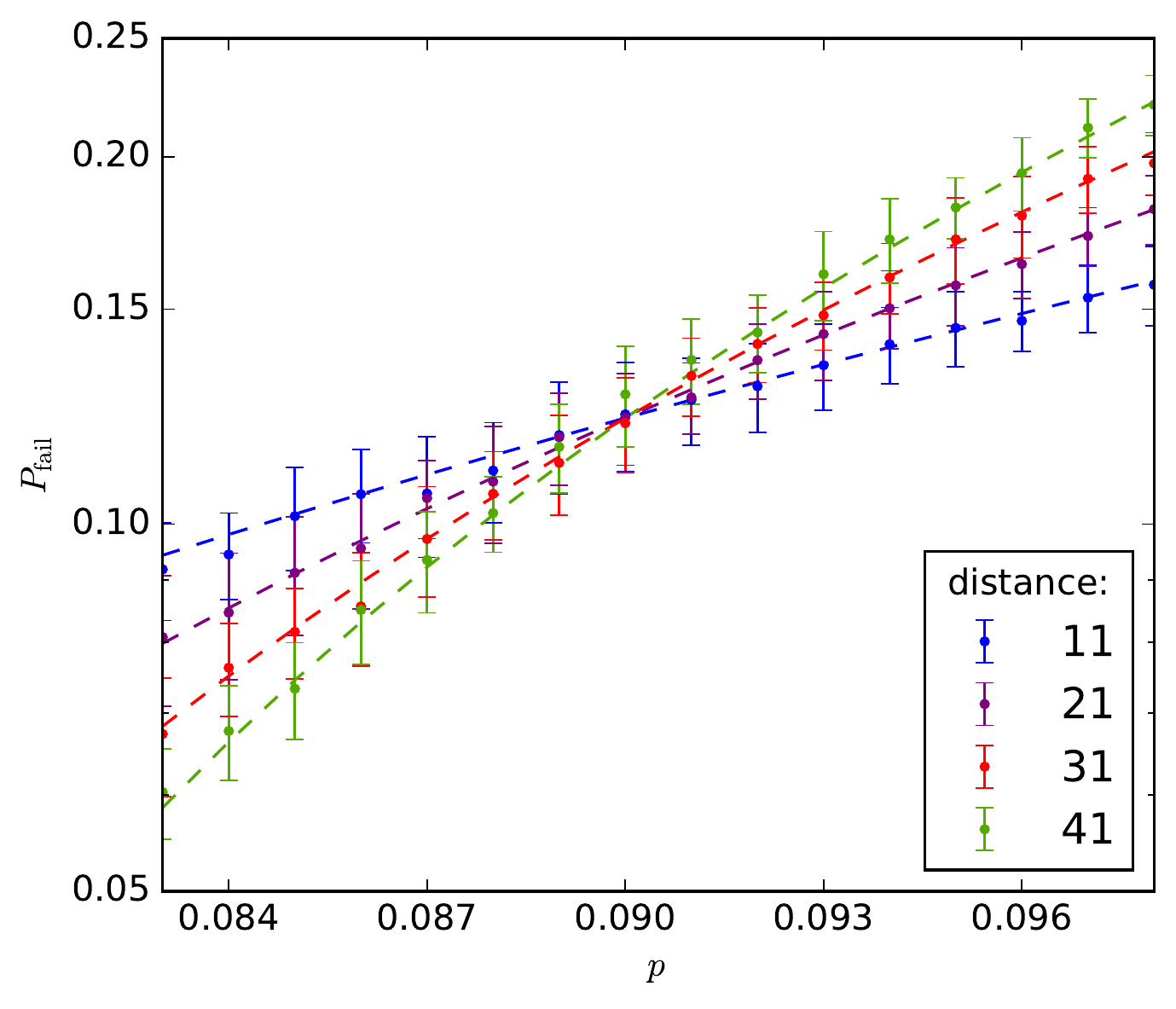}
		\caption{\label{Fig:ErrorThreshold} Threshold plot for an independent and identically distributed bit-flip noise model. The logical failure rate $P_\textrm{fail}$ is plotted as a function of physical error rate $p$ for codes of different distance $d$. The threshold, indicated by the intersection of the curves, is found to be $9.0\%$. The error bars show the standard deviation of the mean logical failure rate where each data point is collected using $\sim 5 \times 10^4 $ Monte Carlo samples. The data collected is fit to the function $f = A{x}^{2}+ Bx + C$ where $x$ is the rescaled error rate $x = (p-p_c)d^{1/v_0}$. Each dashed line indicates this fitting for a given system size.}
	\end{figure}

	The threshold is the critical physical error rate $p_c$ below which the logical failure rate $P_{\mathrm{fail}}$ can be suppressed given increasing system sizes. The threshold is indicated by the intersection of curves recorded for different system sizes as seen in Fig.~\ref{Fig:ErrorThreshold}. We calculate that
	$$p_c = 9.0\%$$ 
	using Monte Carlo simulations, with 50 000 samples having been collected for each data point.

	We fit data close to the crossing to a Taylor expansion truncated to the quadratic term $f = A{x}^{2}+ Bx + C$, where the function $f$ is expressed in terms of the rescaled error rate $x = (p-p_c)d^{1/v_0}$. This method has been explained in greater detail in Ref.~\cite{Wang03}. We obtain the value of the critical exponent $v_0 = 1.422$ and the constants $A=1.215$, $B=0.783$ and $C=0.122$.	

The threshold we find demonstrates a modest improvement over the value $\sim 8.7\%$ obtained with other matching decoders for the hexagonal lattice color code~\cite{Delfosse14}. However, we note that our decoder is unable to match the performance of a maximum-likelihood decoder~\cite{Katzgraber09, Katzgraber10, Tuckett19} or even a neural-network decoder~\cite{Maskara19}. It may be interesting to determine the types of errors that lead the decoder to fail for error rates $ 9\%\le p \le 10.9\% $. This information may give us new insights into ways we can improve the M\"obius decoder. In what follows we develop this decoder by finding and comparing two alternative corrections to find a better result at very low error rates, i.e., the comparative decoder. Surprisingly, we found that the threshold we obtained, $\sim 9.0\%$, was insensitive to this improvement.  
	
	\subsection{Exhaustive simulations}
	\label{SubSec:Exhaustive}
	We exhaustively test a comparative decoder over system sizes $d \le 13$ for errors of weight $w\le (d-1)/2$. With the change introduced to our original M\"obius decoder as described below, no logical failures were detected for all system sizes evaluated in this search. At $d = 13$ we test $ \sim 5 \times 10^{9}$ error configurations.

In SubSec.~\ref{SubSec:AltCorrection} we motivated the need to compare different low-weight corrections to make a better decision to recover the code state, and in Appendix~\ref{App:AlternativeCorrections} we gave details on how to obtain an alternative correction. In what follows we describe how we compared the original and alternative corrections to obtain this result. We also discuss other methods of comparing different choices of correction operator.

	\begin{figure}
		\includegraphics[width=0.5\columnwidth]{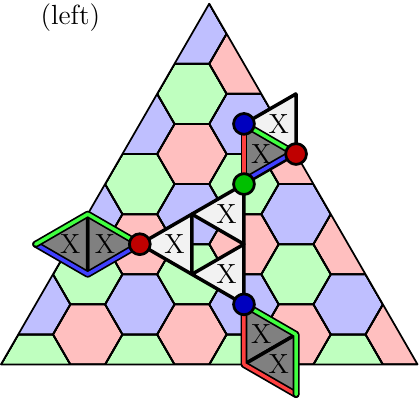}\hfill
		\includegraphics[width=0.5\columnwidth]{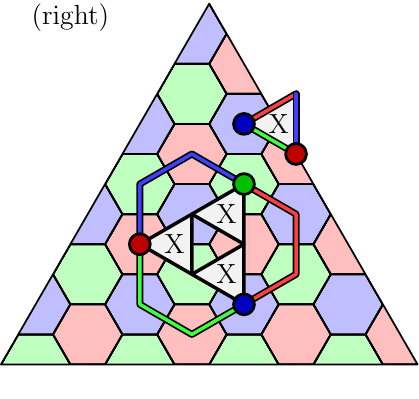}
		\caption{\label{Fig:tricky} We observe an error of weight $w=4$ shown by the qubits in white on the distance $d=9$ lattice. The minimum weight matching decoder will choose the (left) set of edges such that $\ell_{or.} = 11$. However, we note that the error configuration in dark gray traced out by these edges is logically inequivalent to the least-weight solution and is of weight $w=5$. The alternate correction (right), which forces an inequivalent path around the M\"obius Strip, finds a set of edges with $ \ell_{alt.} = 12$ that correctly traces out the true least-weight error. This phenomenon is an instance of the branches we have discussed in Fig.~\ref{Fig:ErrorsAndEdges}.} 
	\end{figure}

To determine whether we should change our correction to the alternative correction, we simply look at the total lengths of the edges returned by the matching carried out to find the two different corrections. We refer the reader back to SubSec.~\ref{SubSec:EdgeWeights} to see how we evaluate the lengths of the edges between defects for a matching. We denote the total lengths of the edges returned by original and alternative matching as $\ell_\textrm{or.}$ and $\ell_\textrm{alt.}$, respectively.
	In this decoder, the logically inequivalent solution replaces the orginal decoder solution if and only if the following two conditions are met:
	\begin{equation}
		\ell_\textrm{alt.} -\ell_\textrm{or.} = \Upsilon,   \label{Eqn:cond1exh}
	\end{equation}
	\begin{equation}
		  (2\ell_\textrm{or.} - d) \textrm{ mod } 4 = 1.  \label{Eqn:cond2exh}
	\end{equation}
and, for all the cases we consider in this exhaustive test, we take $\Upsilon = 1$.

We motivate the conditions given in Eqns.~(\ref{Eqn:cond1exh}) and~(\ref{Eqn:cond2exh}) using the examples we have discussed in SubSec.~\ref{SubSec:AltCorrection} as follows: Eqn.~(\ref{Eqn:cond1exh}) requires that the matching for original and alternate corrections, $\ell_\textrm{or.}$ and $\ell_\textrm{alt.}$, differ in their length by 1. Let us look at a case of concern given by the example in Fig.~\ref{Fig:ErrorsAndEdges}(d). In the example, if a branching error of this type has weight $w = (d-1) / 2$ we argued that the decoder would identify an incorrect matching with $\ell_\text{or.} \sim 2  \times [d - w] = 2 \times [(d+1) / 2] = d+1$, where the decoder matches the defects to the boundaries. On the other hand, we find $\ell_\textrm{alt.} \sim 2\times w  = 2\times [(d-1) / 2 ] +3 = d+2 $. As we see, the difference between $\ell_\textrm{or.}$ and $\ell_\textrm{alt.}$ is $1$, but that the alternative solution is the correct one. For the system sizes of interest in the exhaustive search, we find that all the cases where the alternative correction provides the correct solution, $\ell_\textrm{or.}$ is always one unit smaller than $\ell_\textrm{alt.}$.

Let us also justify Eqn.~(\ref{Eqn:cond2exh}). For $d\le 13$, while the condition given in Eqn.~(\ref{Eqn:cond1exh}) is necessarily satisfied to consider choosing the alternative correction, this criterion is not sufficient to make any such change. A more careful examination shows us that in the cases that the original matching gives the wrong correction for a weight $w= (d-1)/2$ error we find an unphysical solution. We find that we can test the physicality of the error. We state that\footnote{Sketch of proof: A single error introduces an odd parity of edges to the unified lattice on the boundary of the single error where edges are weighted according to SubSec.~\ref{SubSec:EdgeWeights}. Therefore, introducing multiple errors of $w$ bit flips mean that the parity of $\ell$ is equal to the parity of $w$, where cancellations due to the intersection of the boundaries of adjacent errors locally change $\ell$ by an even parity, and thus do not affect the overall parity of $\ell$. The matching may follow a path that does not track the boundary of the error. In which case the path is deformed from the boundary by a trivial cycle on the unified lattice. Given that all trivial cycles have an even length (where we consider that trivial cycles may cross creases and corners), changing the correction by a trivial deformation does not change the parity of $\ell$ where, again, the difference due to the intersection of an edge with a trivial cycle will not change the overall parity of $\ell$.}
$\ell \textrm{ mod } 2  = w \textrm{ mod } 2$.
For close cases then, in the sense of the condition given in Eqn.~(\ref{Eqn:cond1exh}), we can test that the matching corresponds to an error of weight $w = (d-1)/2$. Substituting $w= (d-1) / 2$ into the unphysical solution, $ (\ell - w )\textrm{ mod }2 = 1 $, gives the condition of Eqn.~(\ref{Eqn:cond2exh}) with a simple rearrangement of the expression.

Let us examine two specific error configurations that are representative of the types of errors we encounter in the exhaustive search that are corrected using the conditions we have proposed.  In Fig.~\ref{Fig:tricky} we show an error of weight $w = 4$ on a $d=9$ lattice that is representative of the error shown in Fig.~\ref{Fig:ErrorsAndEdges}(d). The initial use of minimum-weight matching finds a set of edges $\ell_\textrm{or.} = 11$ that trace out a string-like error that is in fact logically inequivalent to the least-weight solution. In contrast, with the comparative decoder, we find an alternative matching that gives a solution with $\ell_\textrm{alt.} = 12$. We check that these values of $\ell_\mathrm{or.}$, $\ell_\mathrm{alt.}$ and $d$ satisfy both Eqns.~(\ref{Eqn:cond1exh}) and~(\ref{Eqn:cond2exh}), thus enabling us to identify the true least-weight error with a branching point.

\begin{figure}
	\includegraphics{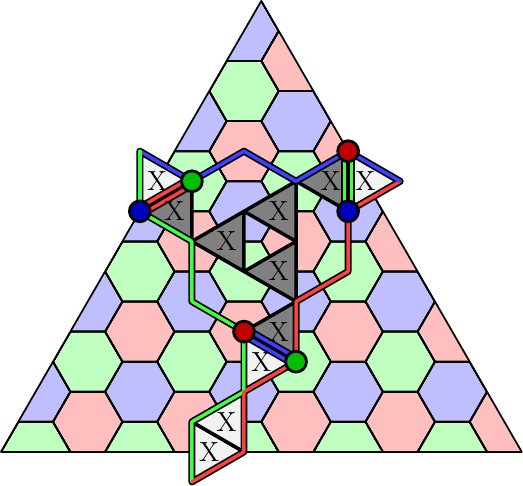}
	\caption{\label{Fig:trickyd11} Error for the $d=11$ color code where the alternative correction gives the least-weight solution with $w = 5$. We have $\ell_\textrm{or.} = 12$ and $\ell_\textrm{alt.} = 13$ where we show the original matching with solid lines and the alternative matching with dashed lines. }
	\end{figure}

At $d=11$, we also find errors where minimum-weight perfect matching misidentifies a correction whereby a high-weight branching error demonstrates a matching with a disproportionately low value of $\ell_\textrm{or.} = 12$. See Fig.~\ref{Fig:trickyd11}. This error is representative of branching errors similar to those in Fig.~\ref{Fig:ErrorsAndEdges}(e) where a high-weight branch gives rise to a low-weight solution to the original matching subroutine. Once again, we have that $\ell_\textrm{alt.} = 13$. Here, the original correction isolates a branch-like error and the alternate correction reflects a string-like error. Once again, we find that these values for $\ell_\textrm{or.}$, $\ell_\textrm{alt.}$ and $d$ satisfy our conditions that indicate we should choose the alternative correction to find a lower-weight solution.

Thus, in conjunction with the observations we have previously stated, under the condition of Eqn.~(\ref{Eqn:cond1exh}) constraining the corrections to arise from errors such that $w_\textrm{or.} + w_\textrm{alt.} = d$ and $ w_\textrm{or.} - w_\textrm{alt.}  = \pm 1$, Eqn.~(\ref{Eqn:cond2exh}) is a simple odd-even rule that ensures that the switch to the alternate correction is only made if the original correction derives from the higher weight $w = (d-1)/2 + 1$ error between the two.

It will be valuable to continue to test these conditions for larger system sizes and finite error rates. We do not expect that the conditions we have used here will continue to be successful for codes of larger distance. Indeed, due to Eqn.~(\ref{Eqn:Branch}), and the discussion around it, we can expect that there will be a larger discrepancy between $\ell_{\textrm{or.}}$ and $\ell_\textrm{alt.}$ than that we check for the condition given in Eqn.~(\ref{Eqn:cond1exh}). Indeed, based on our discussion in SubSec.~\ref{SubSec:AltCorrection}, we conjecture that a general condition will have $\Upsilon $  scaling like $d/14 \sim d / 2 - 3d / 7$. However, we do not observe error configurations that give rise to even lower matching lengths, as we might expect for errors such as those in Fig.~\ref{Fig:ErrorsAndEdges}(e), justifying our choice of $\Upsilon = 1$ for the system sizes we test. Furthermore, it will be interesting if we can generalise the condition in Eqn.~(\ref{Eqn:cond2exh}) for errors of weight greater than $w = (d-1)/2$ and to determine if, in fact, there are additional conditions that can help us determine if we have found the correct solution following the initial matching subroutine. Discovering more general conditions will require a deeper understanding of the geometrical intricacies of the color code lattice.

To continue this analysis, we will need better methods for testing errors at low error rates. At $d = 15$ there are over half a trillion errors of weight $(d-1)/2$; over one-hunrded times more samples than we have tested at $d=13$. As such this test is impractical to run, even with a high-performance computing cluster. We might consider other diagnostics to test the performance of different conditions at larger system sizes, for instance, the methods proposed in~\cite{Bravyi13}. Perhaps we can even find analytic expressions that explain the performance of the decoder that can be reached.

\subsection{Estimating the weight of a least-weight correction}

To generalise our result it will be interesting to find better ways of estimating the weight of a correction for some given matching, in order to accurately compare the results of the inequivalent matching subroutines. Let us remark on some other methods we have tested by exhaustive search to find a decoder that can correct all errors of weight $(d-1) /2$. The tests we are about to describe were carried out for system sizes $d\le 11$. 

Refs.~\cite{Delfosse14} and~\cite{Kubica19} have both considered methods of estimating the weight of a correction given the output of matching subroutines on the different restricted lattices of the color code. Roughly speaking, Ref.~\cite{Delfosse14} explains that a correction operator is supported on all of the qubits inside the boundary marked by a set of edges returned from the matching on all three restricted lattices. Ref.~\cite{Kubica19} builds on this idea, and shows that we can find a correction that is proportional to the lengths of the edges returned from two matching subroutines from two of the three different restricted lattices, up to some local corrections.

\begin{figure}
		\includegraphics[width=0.5\columnwidth]{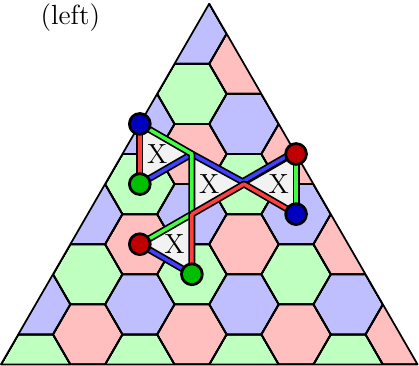}\hfill
		\includegraphics[width=0.5\columnwidth]{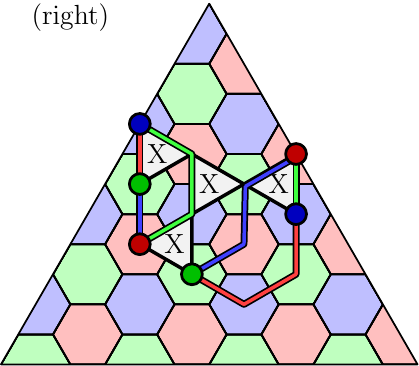}
		\caption{\label{Fig:tricky2} A weight $(d-1)/2 = 4$ error on a $d=9$ color code. A correction of weight five can introduce a logical error. As such it is important to determine that the weight of the correction associated to this matching is $w=4$. A basic implementation of a matching subroutine is equally likely to choose between the (left) and (right) output, as both have $  \ell = 12$. However, only the matching and choice of paths on the matching of (left) will predict a correction of weight four using the methods of Refs.~\cite{Delfosse14, Kubica19}. For the matching given on the left-hand side we see that the edges bound the low-weight error exactly, thereby showing that the correct weight will be obtained using the method of Ref.~\cite{Delfosse14}, for instance. Likewise, the set of edges, (left), will give the correct weight using the methods of Ref.~\cite{Kubica19}. Both of these methods will overestimate the weight of the correction given the matching shown on the opposite lattice,~(right).} 
	\end{figure}

We found with a basic implementation of minimum-weight perfect matching that we were not able to correct all weight $(d-1) / 2$ errors using either of the methods in Refs.~\cite{Delfosse14,Kubica19} to estimate the weight of a correction for a given matching. A problem we encountered is that there are multiple solutions to the matching problem for a given pattern of defects. We give an example in Fig.~\ref{Fig:tricky2} together with an explanation in the figure caption. Roughly speaking, while some solutions to minimum-weight perfect matching enable us to evaluate the correct weight for the least-weight correction, other degenerate solutions lead the decoder to overestimate the weight of the correction which can cause a comparative decoder to fail.

We add that we considered a number of ways to improve the solution returned by the minimum-weight perfect-matching algorithm  to steer the matching subroutine to a favourable configuration to find the correct weight for the correction using the methods proposed in Refs.~\cite{Delfosse14, Kubica19}. Specifically, we considered adding small adjustments to the edge weights for the input graph to bias the output to a more appropriate set of edges, and we were also selective of the paths we chose for the edges of the returned matching. We also tried comparing different subsets of edges according to their colors to find lower weight corrections with the method proposed in~\cite{Kubica19}. However, we were unable to find a strategy that would consistently correct all weight $(d-1) / 2$ errors. 

It is interesting that the diagnostic we found to find the correct solution depends on global features of the two alternative solutions found by different matching subroutines. In contrast, the methods in Refs.~\cite{Delfosse14, Kubica19} are able to locally estimate the weight of a small error cluster. In future work it will be interesting to learn how to generalise the conditions we have found to determine the best correction between that obtained with the original and alternative matching to go beyond the special case we have considered already with errors of weight $(d-1)/2$ and codes of small distance. We may find that the global diagnostics we have proposed that take three integer values; $\ell_\textrm{or.}$, $\ell_\textrm{alt.}$ and $d$, can be combined with local methods for evaluating the weight of an error.

	\section{Discussion}

	In summary, we have shown how we can reduce logical failure rates for the color code with a matching decoder. We have demonstrated two innovations to correct high-weight errors, namely, we introduced a unified lattice for matching, which manifests as a M\"obius strip for the triangular code, and we have also developed a method to arrive at lower-weight corrections using additional matching subroutines that produces inequivalent solutions to the decoding problem. To demonstrate the performance of our decoder we have evaluated logical failure rates at low $p$, and we have also conducted exhaustive analyses. Notably, our results generalise readily to the fault-tolerant setting where measurements are unreliable. This will be important to develop practical decoders for use with real physical hardware.

	Our analysis suggests that there are ways to optimise our decoding strategy further. It may be fruitful to explore avenues by which soft information~\cite{pattison2021improved} can be integrated into the correction procedure via message passing in order to improve its performance. Let us also remark that there are now a number of different approaches to decode the color code, including approximate maximum-likelihood decoders~\cite{Tuckett19}. It may be interesting to find new ways of comparing our decoder side by side with an optimal decoder to determine the classes of errors that are potentially correctable where our decoder is currently failing.

	Ultimately, it may be valuable to generalise the matching decoder to more general classes of codes. The notion of stabilizer group symmetries offers us a natural route towards one such generalisation. A question that arises when discussing the use of symmetries to find matching decoders is how in general should we choose to over complete the generating set of the stabilizer group to make useful symmetries for decoding. For topological codes we typically use physical insights to find the symmetries that give rise to high-performance matching decoders. Specifically, the stabilizer relations for topological codes correspond to conservation laws among the low-energy excitations of their underlying phase~\cite{Kitaev03, Brown20parallelized}. Nevertheless, the example we have presented has shown us that we should even reexamine the symmetries we use to implement decoders for topological codes. To this end, in Appendix~\ref{Sec:Generalisations}, we discuss a number of ways that the methods we have developed here generalise to other decoding problems associated to the color code, and its variants. Our generalisations include a discussion on alternative symmetries that may be useful for decoding depolarising noise, and for fault-tolerant error correction, where we make use of the connection between the color code and the three-fermion model~\cite{Rowell09, Kesselring18, Roberts20}. We also discuss how we can improve error correction with Majorana surface codes and higher-dimensional color codes by examining the symmetries of the boundaries of these codes. Finally, we discuss how these methods apply to single-shot error correction with the gauge color code. It is our hope that the ideas and tools we have developed here may inspire generalisations of the minimum-weight perfect-matching decoder to other codes in the future.

	\begin{acknowledgements}
		We thank S. Bartlett, N. Breuckmann, N. Delfosse, S. Flammia, M. Kesselring, A. Kubica, G. Nixon, S. Roberts, D. Tuckett and M. Vasmer for helpful and encouraging conversations. We are particularly grateful to M. Newman for showing us the low-weight errors that inspired this work, to C. Jones for showing us some challenges associated to correcting branching errors, to P. Bonilla Ataides for conversations and for critically reading our manuscript, and to A. Doherty for pushing us to complete this project. BJB is also thankful for a great number of ideas and insights, as well as encouragement, over many conversations with David Poulin.  KS is grateful for the hospitality of the School of Physics and the Quantum Theory Group at the University of Sydney. This work is supported by the Australian Research Council via the Centre of Excellence in Engineered Quantum Systems (EQUS) project number CE170100009. BJB also received support from the University of Sydney Fellowship Programme. The authors acknowledge the facilities of the Sydney Informatics Hub at the University of Sydney and, in particular, access to the high performance computing facility Artemis.
	\end{acknowledgements}

	\appendix

	\section{Interpreting edges from the matching subroutine}
	\label{App:SymmetriesAndLogicals}

	Here we explain why an error that anti commutes with the logical operator shown in Fig.~\ref{Fig:ColorCodeSymmetry} necessarily produces an odd parity of defects on either side of the logical operator of interest. To do so, let us divide the symmetry $\Sigma_\mathbf{r} = \Sigma_\mathbf{r} ^A \cup \Sigma_\mathbf{r} ^B$ into two disjoint subsets where $\Sigma_\mathbf{r} ^A $($ \Sigma_\mathbf{r} ^B$) is the subset of faces above(below) the horizontal red line in Fig.~\ref{Fig:ColorCodeSymmetry}.

\begin{figure}
	\includegraphics{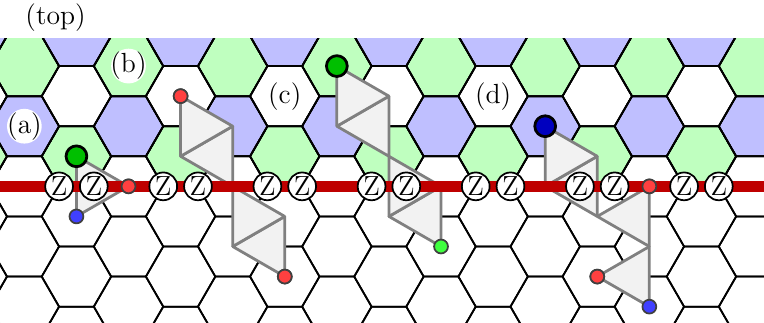}
	
	\smallskip
	
	\includegraphics{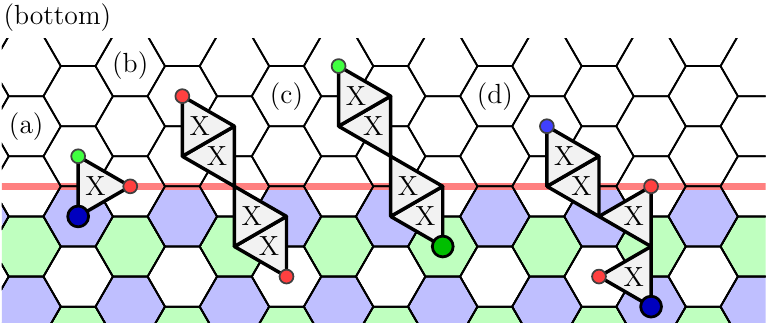}
	\caption{\label{Fig:ColorCodeSymmetries}  (top)~We show the stabilizers of $\Sigma_\mathbf{r}^A$ where we bring the logical operator $\overline{L}$ to the foreground. In the figure we have that $\prod_{A} s \sim \overline{L}$ where we take the product over $s \in \Sigma_\mathbf{r}^A$. All errors~(a),~(c) and~(d) anti commute with $\overline{L}$. As such, for consistency, they necessarily create an odd parity of defects on the faces of $\Sigma^A_\mathbf{r}$. Error~(b) commutes with $\overline{L}$ as it has no common support with the logical operator. It therefore creates an even parity of defects on the faces of $\Sigma_\mathbf{r}^A$. (bottom)~For completeness we show the faces of $\Sigma^B_\mathbf{r}$ together with the errors in the foreground. Again, errors~(a),~(c) and~(d) create a single defect on the faces of $\Sigma^B_\mathbf{r}$ due to their commutation relation with the logical operator $  \overline{L} \sim \prod_B s $. Error~(b) creates no defects on the faces of $\Sigma^B_\mathbf{r}$.  }
\end{figure}

	Let us examine the subset $\Sigma_\mathbf{r} ^A $ in Fig.~\ref{Fig:ColorCodeSymmetries}(top). In the picture, we bring the logical operator $\overline{L}$ supported on the red line to the foreground of the image. The figure also illustrates that 
	\begin{equation}
		\prod_{s \in \Sigma_\mathbf{r} ^A } s \sim \overline{L}, \label{Eqn:SymmetryLogical}
	\end{equation}
	where we use a $ \sim $ symbol to indicate that we are only interested in the support of the operator $\prod_{s \in \Sigma_\mathbf{r} ^A } s$ that is on the restriction of the lattice shown in the figure panel. To be more precise we have implicitly assumed that we have stabilizers, $\prod_{s \in \Sigma_\mathbf{r} ^A } s$ and $\prod_{s \in \Sigma_\mathbf{r} ^B } s$, that can clean~\cite{Bravyi09} a logical operator $\overline{L}$ far away from its own support.

	We can use Eqn.~(\ref{Eqn:SymmetryLogical}) to determine the commutator between $E$ and $\overline{L}$. Indeed, the relationship in Eqn.~(\ref{Eqn:SymmetryLogical}) shows us that errors that anti commute with $\overline{L}$ must give rise to an odd number of defects on the faces of $ \Sigma_\mathbf{r} ^A $. Operators~(a),~(c) and~(d) are examples of errors that anti commute with the logical operator. As expected, all three of these errors give rise to a single defect on $ \Sigma_\mathbf{r} ^A $. The figure also shows that the support of these errors overlaps with the support of the logical operator at a single site. In contrast, error~(b) in the figure has no common support with the logical operator. Neither does it produce any defects on $ \Sigma_\mathbf{r} ^A $, as such, this error is also consistent with Eqn.~(\ref{Eqn:SymmetryLogical}). For completeness, we show the same errors for a third time in Fig.~\ref{Fig:ColorCodeSymmetries}(bottom), with their defects now shown only on the subset of faces $\Sigma_\mathbf{r}^B$. Again, errors~(a),~(c) and~(d) all create a single defect on $\Sigma_\mathbf{r}^B$ whereas the error at~(b) creates no defects on this subset. This is necessarily true given the definition of $\Sigma_\mathbf{r}$ and $\Sigma_\mathbf{r}^A$.

The method we have given to find a correction operator is presented in a way that is readily generalised to other stabilizer codes. We have assumed that we can find an operator that recovers a code state $C'$, and that we can find suitable elements of the stabilizer group that clean the logical operator $\overline{L}$ far away from its support. It will be interesting to learn if minimum-weight perfect-matching decoders will give a good performance for other codes where, perhaps, single-qubit errors give rise to a large number of defects with respect to some well-chosen symmetry. Examples of decoding algorithms for stabilizer codes beyond two-dimensional topological codes are presented in Refs.~\cite{Brown20parallelized, Nixon21}. Further work is required to determine the success of minimum-weight perfect-matching decoders in the general case.

	\section{Using matching to find a likely error}
	
	\label{App:MWPM}
	
	Here we justify weighting edges connecting two defects on the restricted lattice according to their separation. We propose a straight forward way of calculating this separation on the hexagonal lattice in Appendix~\ref{App:MeasuringEdgeWeights}.

	The negative logarithm of the probability that the independent and identically distributed noise model caused an error should be proportional to the weight of the error. We obtain a lower bound on the weight of the error by approximating the error as a series of strings that connect pairs of defects on the restricted lattice. Let us express an error as a product of strings, i.e.,
	\begin{equation}
		E = \prod_{\alpha \in\mathcal{A}}  \alpha , \label{Eqn:Strings}
	\end{equation}
	where $\alpha = \prod_{v \in e(\alpha)} X_v$ are string-like operators that create an even number of defects at their endpoints with respect to a symmetry and $\mathcal{A}$ denotes the set of string-like operators that produced the error syndrome. We show examples of string-like errors that make up $\mathcal{A}$ in Fig.~\ref{Fig:ColorCodeSymmetry}(a),~(c) and~(d).

	Let us assume that bit-flip errors occur with probability $p$ such that the probability that error $E$ occurs is
	\begin{equation}
		\textrm{prob}(E) = (1-p)^n \left( \frac{p}{1-p} \right)^{|E|}, \label{Eqn:LogProb}
	\end{equation}
	where $|E|$ denotes the weight of $E$. 
	With the definition in Eqn.~(\ref{Eqn:Strings}) in place, we can write the negative logarithm of probability that error $E$ occurred as
	\begin{equation}
		\log \textrm{prob}(E) \ge   \log (1-p)^n  + \log \left( \frac{p}{1-p} \right) \sum_{\alpha \in \mathcal{A}} \textrm{length}(\alpha) ,
	\end{equation}
	where $\textrm{length}(\alpha)$ measures the separation between the defects created by $\alpha$ such that $|E| \ge \sum_{\alpha \in \mathcal{A}} \textrm{length}(\alpha)$.

	The decoder looks for an error $E'$ such that $E'$ has a syndrome that is consistent with that produced by $E$ and where $ | E' |$ is minimal as this will correspond to the most probable error that caused the syndrome according to Eqn.~(\ref{Eqn:LogProb}). To estimate a least weight correction with respect to a symmetry we look for a minimal solution 
	\begin{equation}
		E' = \sum_{\alpha' \in \mathcal{A}'} \alpha ' ,
	\end{equation}
	where $\mathcal{A}'$ are the set of strings that give rise to the error $E'$ whose error syndrome is equal to that of error $E$.
	
	We use the minimum-weight perfect-matching algorithm to find an error that produced the error syndrome with high probability. Given an independent and identically distributed noise model where bit flips occur at a low rate, we look for a low-weight correction operator. We find a low-weight correction by using minimum-weight perfect matching where the defects of a symmetry of the code are vertices of the input graph and we weight edges according to their separation. The edges of the matching returned from the algorithm correspond to strings $\alpha’$ of $E’$.

	\section{Measuring the weights of edges on a restricted lattice}
	\label{App:MeasuringEdgeWeights}
	
\begin{figure}
		
		\includegraphics{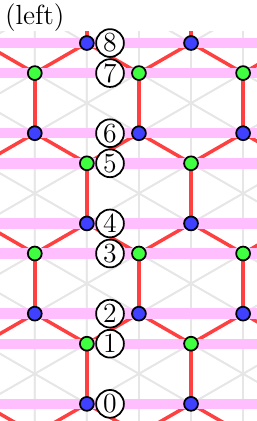} \, \includegraphics{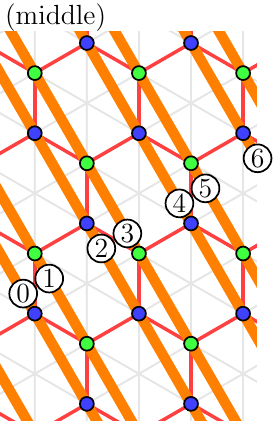}  \, \includegraphics{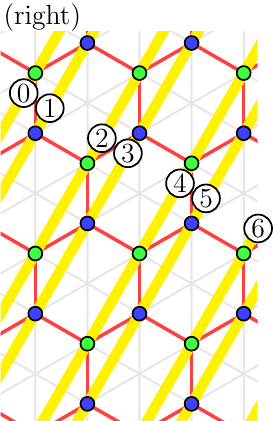}

		\smallskip
		
		\includegraphics{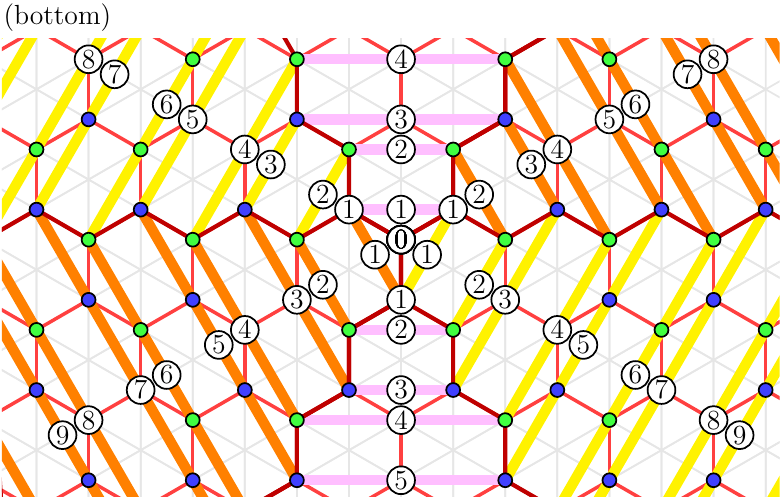}

		\caption{Evaluating the distance between two vertices on the hexagonal lattice. (left),~(middle) and~(right)~show three non-orthogonal axes we use to find the distance between two vertices on the hexagonal lattice. We label the axes shown on the left, middle and right figure $x_0$, $x_2$ and $x_4$, respectively. The brightly colored lines in each of these figures show all the vertices with a common value according to this coordinate system. (bottom)~We show the distance of each of the vertices from the central vertex marked with a $0$ with bold contour lines overlaying the red hexagonal lattice. \label{Fig:Metric}}
	\end{figure}

	Here we show how to evaluate the separation between two defects on the hexagonal restricted lattice. Let us give each vertex a coordinate with three integer values $x = (x_0, x_2, x_4)$. We show them in Figs.~\ref{Fig:Metric}(left),~(middle) and~(right), respectively, where vertices with a common coordinate value lie on a bold line overlaying the lattice.  We call them $x_0$, $x_2$ and $x_4$ as they, respectively, align along the perpendicular to the clock-hand that points along twelve-hundred hours, 2 o-clock and 4 o-clock.

	Given this coordinate system it is easy to find the separation between any two defects. Suppose we have two defects at locations $x = (x_0, x_2, x_4)$ and $y = (y_0, y_2, y_4)$, we find that the smallest number of edges $w$ between $x$ and $y$ is obtained by the formula
	\begin{equation}
		w = \max ( \Delta_0, \Delta_2, \Delta_4),
	\end{equation}
	where $\Delta_j = |y_j - x_j|$. In Fig.~\ref{Fig:Metric}(bottom) we show contours marking the number of edges a given vertex lies from the central vertex denoted $0$. Lastly, let us remark that this coordinate system allows us to count the number of paths of length $w$ between two vertices along the hexagonal lattice. We simply state the result and leave it to the reader to check its validity. We find that 
	\begin{equation}
		\#\text{paths}(x,y) = \frac{r! }{ s!(r-s)!},
	\end{equation}
	with 
	\begin{equation}
		r = 2 \Delta_\textrm{max.} - \Delta_\text{med.} - \Delta_\textrm{min.} , \quad s = \Delta_\textrm{max.} - \Delta_\textrm{med.}, 
	\end{equation} 
	where $\Delta_{\textrm{max.}} \ge \Delta_{\textrm{med.}} \ge \Delta_\textrm{min.}$ are the maximum, median and minimum values of the set $\Delta_0$, $\Delta_2$ and $\Delta_4$, respectively. One can use this formula to determine the degeneracy of a string-like error~\cite{Stace10}. We note that we do not include this term in the edge-weight function for our implementation of the decoder where we match defects on the unified lattice. Indeed, this expression would require modification to deal with edges that pass over creases on the M\"obius strip. We expect that including an adaptation of this term to the edge weights of the input graph to the matching subroutine may improve the performance of the decoder.

	\section{Finding an alternative correction}
	
	\label{App:AlternativeCorrections}

\begin{figure}
		\includegraphics{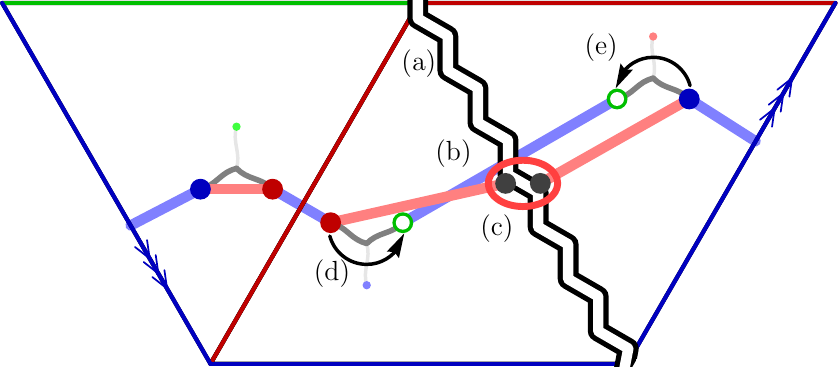}
		\caption{\label{Fig:Dummies} Finding an alternative correction using an additional matching subroutine. The initial matching is shown by blue edges. We modify the input graph to the second subroutine to find an alternative correction. We introduce a `tear' to the M\"obius strip, see the zig-zag line at (a). In the second subroutine, no defects can pair using an edge that crosses the tear.  We also remove all pairs of vertices from the initial matching that are paired by an edge that crosses the tear. For example, see the pair of defects connected by the dashed blue edge,~(b). To find an inequivalent matching we introduce a single dummy node on either side of the tear,~(c). The inclusion of these dummy nodes force the the matching to find an alternative correction, shown by the red edges. The two defects that are matched by the two dummy nodes form a new edge, see the endpoints of the edge at~(d) and~(e). Finally, we try to reduce the weight of the edge found using the dummy vertices by looking for a shorter route back over the tear using the edges that were removed when we made the tear. The defects found by the dummy nodes are subsequently paired to the green defects, connected by the dashed line, that were initially removed to obtain the alternative correction when we made the tear. The last step recovers the matching shown in Fig.~\ref{Fig:TwoCorrections} in the main text.}
	\end{figure}

	Here we explain how we use minimum-weight perfect matching to find a second low-weight correction that is logically inequivalent to the first matching. The idea is summarised in Fig.~\ref{Fig:Dummies}, and is developed from an idea proposed in Ref.~\cite{HutterWoottonLoss13}. The goal of this exercise is to find a second low-weight correction that is close to the weight of the correction obtained by the initial matching, except where the second correction differs from the first correction by a logical operator. We therefore seek a matching that differs from the first by a homologically non-trivial cycle about the M\"obius strip. Here we explain how to use minimum-weight perfect matching to find such a correction.

	We perform matching on an alternative manifold where we introduce a `tear' to the M\"obius strip, see Fig.~\ref{Fig:Dummies}(a). Specifically, the tear represents a barrier that prevents any defects from pairing across the tear. As we explain in the caption of Fig.~\ref{Fig:Dummies}, we remove all pairs of defects that were paired by an edge that crosses the tear in the initial matching subroutine, see Fig.~\ref{Fig:Dummies}(b). To find an alternative correction we introduce a single pair of dummy nodes to the matching graph that begin on either side of the tear, see Fig.~\ref{Fig:Dummies}(c). The dummy defects may only pair by an edge with a very high weight that wraps around the manifold, as such, they find an alternative route via the other defects on the lattice.
	
	The two defects that are paired with the dummy defects form a single new edge that crosses the tear. We show the end points of the new edge found by the dummy defects at Fig.~\ref{Fig:Dummies}(d) and~(e). With this, we see why it is necessary to remove the defects that were previously paired by edges that cross the tear. If these defects remain on the torn manifold, the dummy nodes might be inclined to pair to these defects to find a second matching that gives a correction that is equivalent to the first. The figure clearly shows that the inclusion of the dummy nodes and the tear force the matching subroutine to propose a correction, shown in red, that is inequivalent to the original matching, shown in blue in the figure.

	Once we have found an alternative correction, we attempt to reduce the weight of the matching using the edges we removed when we introduced the tear. Specifically, we attempt to find a lower-weight correction by connecting the two vertices that were paired via the two dummy nodes, let us call them $u_l$ and $u_r$, to the defects that were removed for the second matching. We show this process in Figs.~\ref{Fig:Dummies}(d) and~(e). To execute this operation computationally, we take a list of edges $e_j$ that we remove from the torn lattice because they cross the tear. These edges $e_j = (l_j,r_j)$ connect vertices $l_j$ and $r_j$ that, respectively, sit to the left and right of the tear. We evaluate $\lambda = \min_j (|u_l - l_j| + | u_r - r_j |)$, i.e., the sum of the lengths of the shortest two edges that connect $u_l$ and $u_r$ via an edge that was removed. Then, we compare $\lambda$ to the weights of edges $|u_l - u_r|$ and $| l_j - r_j |$ to determine if we replace edges $u_l - u_r$ and $l_j - r_j$ with edges $u_l - l_j $ and $ u_r - r_j $ to find a lower weight correction.

Finally, let us comment on the locations of the dummy defects. There are $d$ locations along the tear where we can place the pair of dummy defects. It is not obvious, {\it a priori}, which location will give the least-weight alternative matching. As such, we perform matching to find the least-weight matching $d$ times, where the pair of adjacent dummy nodes are translated along the lattice sites of the tear with each variation of the subroutine. Once we have performed all of the subroutines we compare the results of all of the subroutines and choose the alternative matching with the least weight. In practice we can perform all $d$ of these matching subroutines in parallel, provided we have already obtained the results of the initial matching subroutine.

\section{Fitting at low error rates}
\label{App:Fitting}

	\begin{figure}
		\includegraphics[width=\columnwidth]{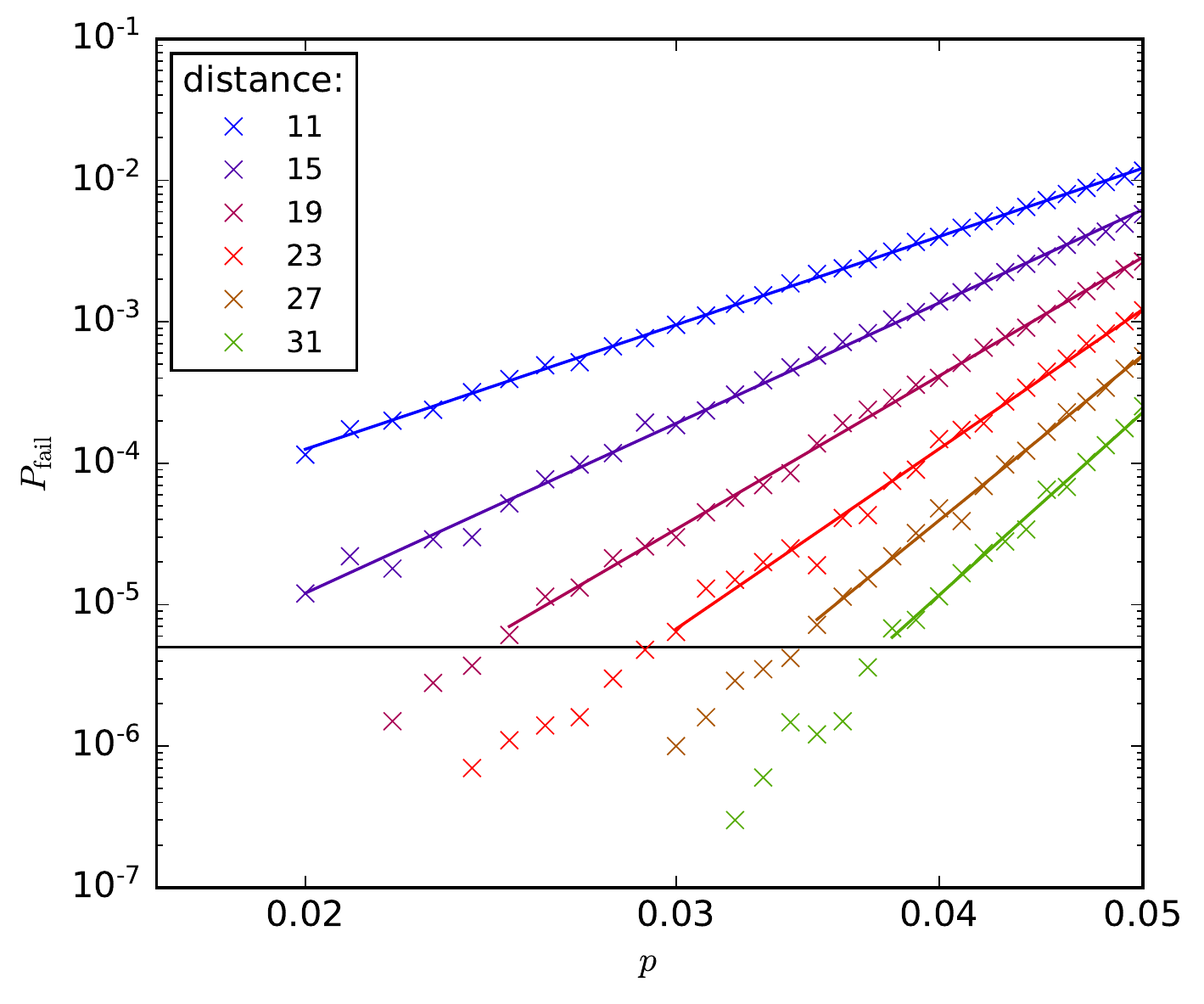}
		\includegraphics[width=0.5\columnwidth]{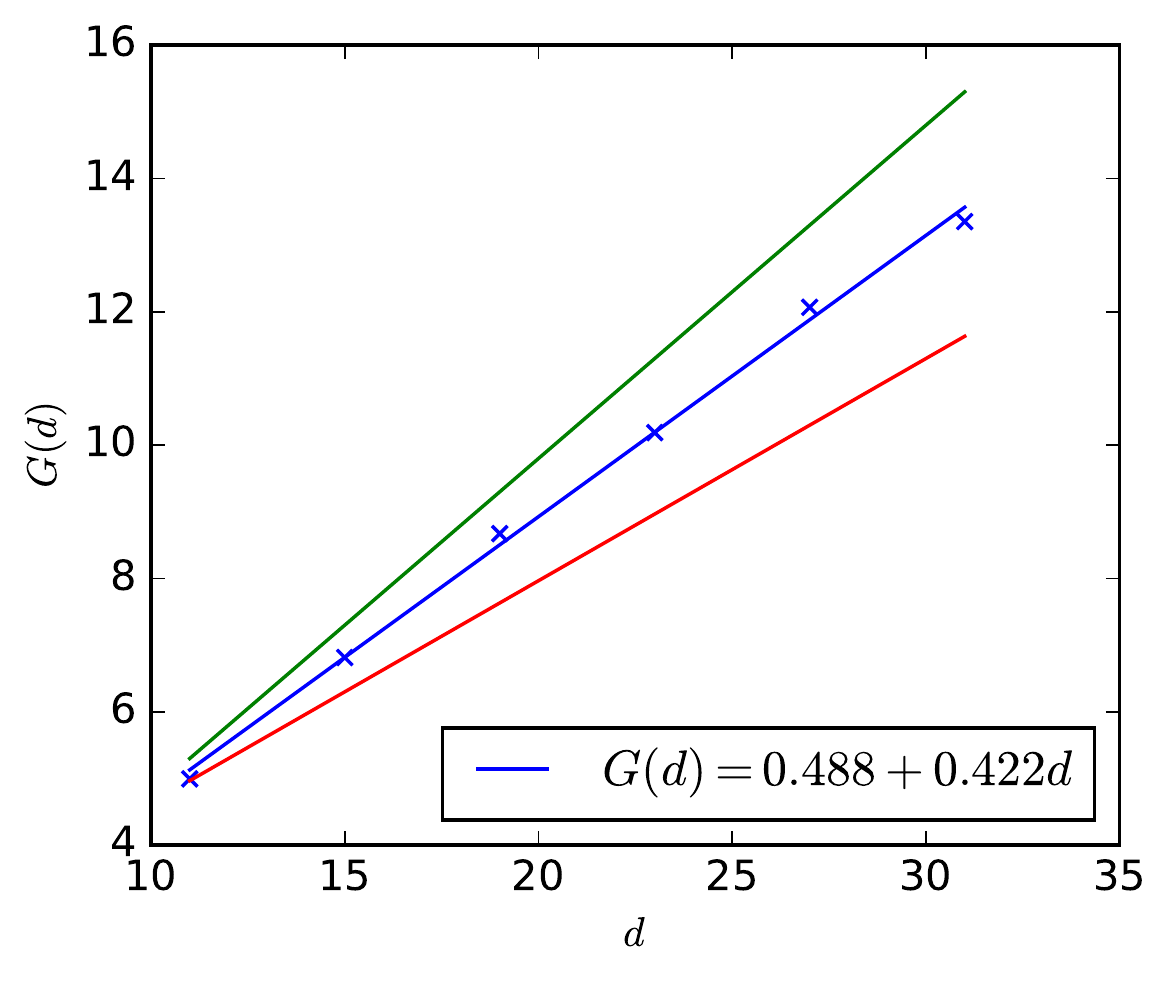}\hfill
		\includegraphics[width=0.5\columnwidth]{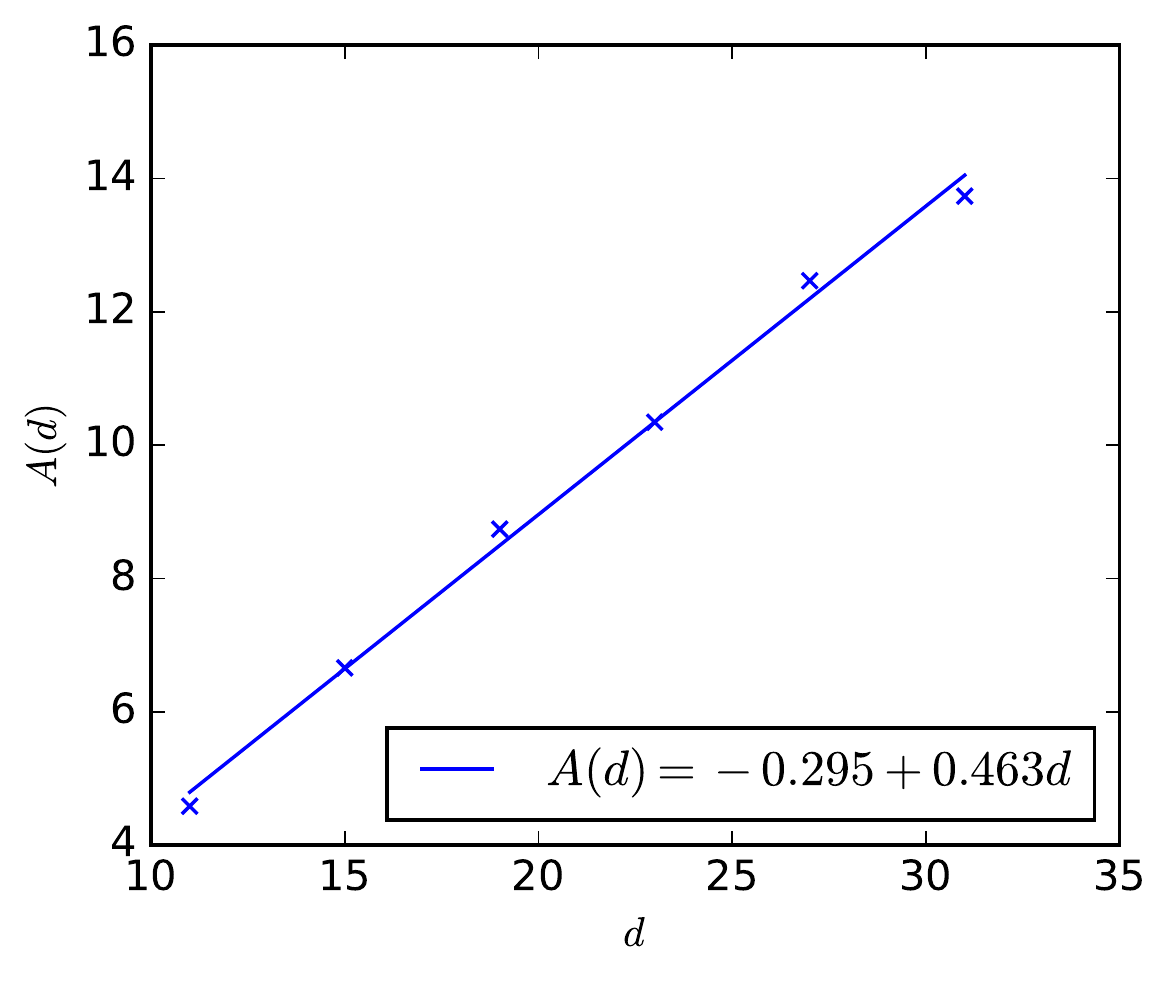}
		\caption{\label{Fig:lowP} (top)~The logarithm of the logical failure rate $P_{\text{fail}}$ is plotted against $\log(p)$ for different system sizes. Each individual plot is fitted to a linear equation, discarding points for which $P_{\text{fail}}$ is lower than $5\times 10^{-4}\%$. These data points sit below the horizontal black line. We then extract the gradients $G(d)$ and intercepts $A(d)$ for each linear fit. (bottom)~The gradients and intercepts of the linear fittings observed in Fig.~\ref{Fig:lowP} are plotted as function of the system size $d$. (bottom left)~For the linear plot of $G(d)$ in blue, the slope is the value of $\alpha$ for our decoder and $\gamma$ is the intercept. The green line has a slope of $1/2$, which is the gradient we expect for a decoder that can correct up to its code distance. The red line has a slope of $1/3$. We read $\alpha$ and $\gamma$ from the gradient and intercept of the blue line, respectively. (bottom right)~For the function $A(d)$, the intercept is $\log \beta  + \gamma \log N$ and the slope is $\alpha \log N$. } 
	\end{figure}

Here we explain how we obtained the values for our ansatz in Eqn.~(\ref{Eqn:ansatz}). We take the logarithm of the ansatz and separate out the terms dependant on $\log p$ such that Eqn.~(\ref{Eqn:ansatz}) takes the following form
	\begin{equation}
	\log {P}_{\text{fail}}={G}({d}) \log p + {A}({d}).
	\end{equation}
	where now
	\begin{equation}
	G(d) = \alpha d + \gamma, 
	\end{equation}
	and
	\begin{equation}
	 A(d) = (\alpha  d + \gamma) \log N + \log \beta.
	\end{equation}
We plot the logarithm of the logical failure rate against $\log p$ at different system sizes as in Fig.~\ref{Fig:lowP}. Subsequently, we make a linear fit for each system size $d$ such that we obtain data points for $G(d)$ from the gradient of each value of $d$ and for $A(d)$ from the intercept for each $d$. Data points used to obtain these fittings are collected using between $10^6$ and $10^7$ Monte Carlo samples, and we discard data points where $P_{\text{fail}}$ is lower than $5\times 10^{-4}\%$. 

We use our data points for $G(d)$ and $A(d)$ to find the fitting parameters for our ansatz. Both $G(d)$ and $A(d) $ have a linear form in code distance $d$. We plot $G(d)$ and $A(d)$ as a function of $d$ in the bottom left and bottom right graph in Fig.~\ref{Fig:lowP}, respectively. Using our data we obtain
	\begin{equation}
	G(d) = 0.422d + 0.488, \label{Eqn:gd}
	\end{equation}
	and
	\begin{equation}
	A(d) = 0.463d  -0.295. \label{Eqn:ad}
	\end{equation}
 To find the fitting parameters then we first read off $\alpha \approx 0.422$ and $\gamma \approx 0.488$ from the gradient and intercept for the linear fit to $G(d)$ as a function of $d$. We are then free to obtain the remaining parameters from the linear fit to $A(d)$. We find ${N} \approx 12.49 $ and $\beta \approx 0.148$ using the values of $\alpha$ and $\gamma$ obtained from the $G(d)$ fitting.

	\section{Generalisations}
	\label{Sec:Generalisations}
	
		\begin{figure}
		\includegraphics{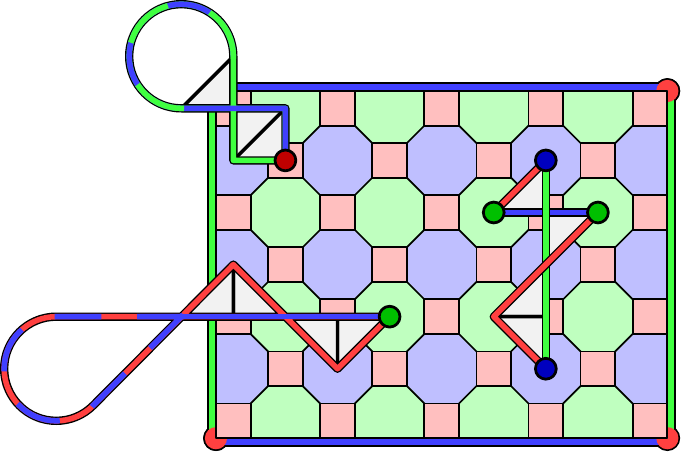} 
		\caption{The color code with two green boundaries and two blue boundaries. All four corners of the lattice are red. Like the color code on the triangular lattice, we can unify the restricted lattices of this code through the boundaries and the corners of the lattice. We show a green defect and a red defect created at the boundary. We show edges that pair them between different restricted lattices via the boundary.
		Furthermore, we can consider a Majorana surface code where an unpaired Majorana mode $\lambda_v$ lies on each vertex and we have stabilizers $S_f = \prod_{v\in \partial f} \lambda_v$. In this picture the red face operators can be regarded as a tetron. Decoding the Majorana fermion surface code is equivalent to correcting bit flips for the color code. \label{Fig:TetronCode}}
	\end{figure}
	
	We have proposed a decoder for the color code on the triangular lattice with three distinct and differently colored boundaries. This case is well motivated due to the capability of this code to perform a complete set of transversal Clifford gates. Nevertheless, we can conceive of other modes of quantum computation that will require alternative color code lattices undergoing more general noise models. Moreover, we may also consider generalising this decoder to higher-dimensional variants of the color code; notable among which is the three-dimensional color code that has a universal transversal gate set when supplemented by gauge fixing. Here we propose ways of generalising the methods we have presented to some other representative variations of the color code.

	\subsection{Alternative boundary configurations and Majorana codes}

\begin{figure}
	\includegraphics{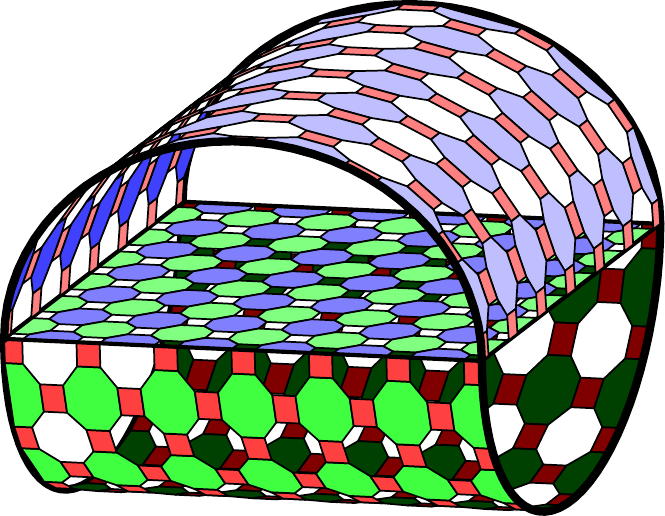}
	\caption{The unified lattice of the color code shown in Fig.~\ref{Fig:TetronCode} is embedded on a manifold with the topology of a torus with a single disk-shaped puncture. \label{Fig:Manifold}}
	\end{figure}

	The unified lattice we have proposed for the triangular lattice can be generalised to other color code lattices with boundaries. In the general case, we can pair defects of different restricted lattices via appropriately colored corners. We show one such alternative lattice in Fig.~\ref{Fig:TetronCode} that encodes two logical qubits. As in SubSec.~\ref{SubSec:Symmetries} we can define a restricted lattice of face operators $\Sigma_\mathbf{u}$ that includes all faces that are not of color $\mathbf{u} = \mathbf{r},\mathbf{g},\mathbf{b}$. Specifically, we can pair defects between the $\Sigma_\mathbf{u}$ and $\Sigma_{\mathbf{v}}$ restricted lattices via boundaries and corners of color $\mathbf{w} \not= \mathbf{u},\mathbf{v}$. Fig.~\ref{Fig:TetronCode} shows two errors that create defects that are paired between two different restricted lattices via the boundaries. We show the topology of the unified lattice in Fig.~\ref{Fig:Manifold}.

Decoding the lattice shown in Fig.~\ref{Fig:TetronCode} may be particularly interesting, since this lattice has already been demonstrated to have a threshold at $\sim 10\%$~\cite{Kubica19} using a restriction decoder. It may be that the threshold will increase further using a unified lattice. Indeed, there may be additional syndrome information that is neglected by the restricted lattice that may improve the performance of the decoder. Moreover, the structure of this lattice is such that it is relatively straight forward to count the number of least weight errors that should lead the color code on this particular lattice to logical failure. As such, studying this model may provide a useful example to evaluate the entropic contributions that affect the performance of a color code decoder.

	We also remark that decoding bit-flip errors on this color code is equivalent to decoding common types of errors acting on a Majorana surface code~\cite{Vijay15, Litinski18, Li18, Viyuela19, Chao20}. In the example shown we can think of the red face operators as the support of a single tetron, i.e., an island where an unpaired Majorana mode lies at each corner of each square face. We therefore see that our methods are readily adapted to decode Majorana surface codes.

	We have argued that it is advantageous to decode the color code using a unified lattice that consists of all of the restricted lattices. We have shown how to combine restricted lattices via the boundaries of a planar color code to form a crease. However, for completeness, we should also consider how we can combine restricted lattices on a continuous lattice with no boundaries. Let us now consider the color code with periodic boundary conditions.

\begin{figure}
		\includegraphics{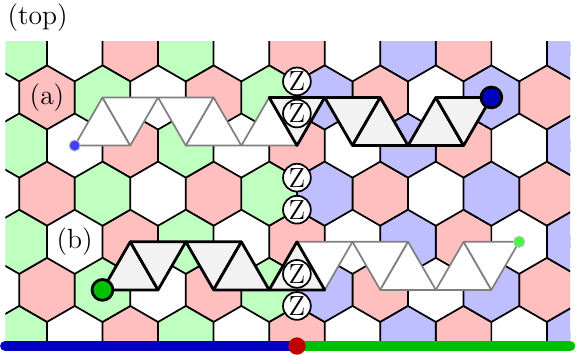}
		\includegraphics{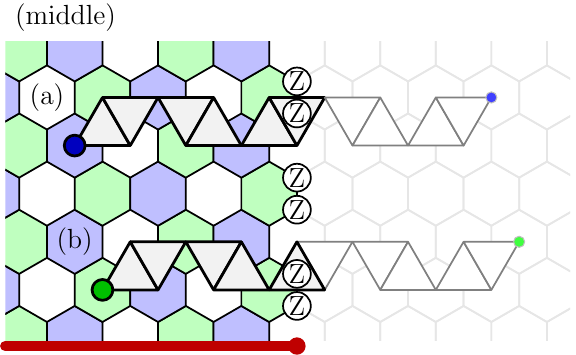}
		\includegraphics{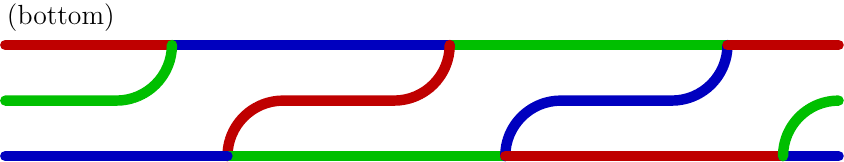}
		\caption{\label{Fig:Seam}A proposal for a unified symmetry for the color code on a toric lattice. (top)~We show an element of the stabilizer group depicted by the product of shaded face operators. It consists of the restricted lattice $\Sigma_\mathbf{b}$ to the left of the lattice and $\Sigma_\mathbf{g}$ to the right of the lattice. The product of these operators is a one-dimensional line of Pauli-Z operators that runs vertically through the middle of the lattice. (middle)~We obtain the same stabilizer term shown in the top figure by taking the product of the green and blue face operators of the restricted lattice $\Sigma_\mathbf{r}$. Together, the set of stabilizers shown in the top figure, and those shown in the middle figure give a symmetry. The union of these two subsets of stabilizers form a seam. (bottom)~We can combine seams around a periodic lattice to find a global symmetry. We propose one suitable configuration by representing the two-dimensional lattices by one-dimensional lines that cut the lattices. We show how these bold lines align with the two-dimensional restricted lattices by the bold lines shown at the bottom of the lattices shown at the top and middle of the figure.}
	\end{figure}

	To this end let us propose a `seam'. This is a continuous line along which all three restricted lattices can be connected. The face operators of the symmetry over a single seam are collectively shown in Figs.~\ref{Fig:Seam}(top) and (middle). In Fig.~\ref{Fig:Seam}(top) we show the restricted lattice corresponding to the symmetry $\Sigma_\mathbf{b}$ to the left of the lattice and $\Sigma_\mathbf{g}$ to the right. We obtain the stabilizer that is the product of Pauli-Z operators along a vertical line at the middle of the figure where the two lattices meet. To recover a trivial operator from the stabilizer group corresponding to a symmetry we combine this operator with that obtained by taking the product of the face operators shown in Fig.~\ref{Fig:Seam}(middle). These are the faces of the $\Sigma_\mathbf{r}$ restricted lattice at the left of the figure. The product of all of these faces gives a symmetry along the seam.
	
	Let us now consider how we correct errors that cross over this seam. We show two errors, Figs.~\ref{Fig:Seam}(a) and~(b) on both the (top) and (middle) image. Let us first consider the error shown at Fig.~\ref{Fig:Seam}(a). The blue defect shown on the $\Sigma_\mathbf{g}$ restricted lattice to the right of Fig.~\ref{Fig:Seam}(top) can be paired to the blue defect on the $\Sigma_\mathbf{r}$ restricted lattice shown in Fig.~\ref{Fig:Seam}(middle) via the seam. The same blue defect does not appear on the $\Sigma_\mathbf{b}$ restricted lattice at the left of Fig.~\ref{Fig:Seam}(top). As such, naturally, we see that this error respects the defect conservation symmetry of the color code.

We also consider the error Fig.~\ref{Fig:Seam}(b). Here the green defect appears to the left of Fig.~\ref{Fig:Seam}(top) and (middle). We can consider pairing this defect to itself via the seam. The green defect does not appear at the right of Fig.~\ref{Fig:Seam}(top) and Fig.~\ref{Fig:Seam}(middle). Once again then, we see that this error respects the defect conservation symmetry, as we expect.

	We have now proposed a seam where three restricted lattices meet. Let us now show how we can combine these lattices over the surface of a torus to make a unified lattice that respects a global symmetry. In Fig.~\ref{Fig:Seam}(bottom) we represent each of the restricted lattices by a one-dimensional line, where these one dimensional lines are shown along the bottom edges of the restricted lattices shown in Figs.~\ref{Fig:Seam}(top) and~(middle) according to the color of the restricted lattice $\Sigma_\mathbf{u}$. They denote the one-dimensional intersection of the restricted lattices along a line that runs orthogonal to the seam. We assume that the seams of the lattice are translationally invariant along the orthogonal direction that is not shown by the one-dimensional representation of the symmetry. Fig.~\ref{Fig:Seam}(bottom) therefore shows how we can consistently combine the restricted lattices about a surface with periodic boundary conditions. Given this unified lattice that combines restricted lattices via multiple seams, we can adapt the method discussed in SubSec.~\ref{SubSec:Symmetries} to find the commutator of an error with some choice of logical operators using minimum-weight perfect matching.

	\subsection{Depolarising noise and fermionic symmetries}
	
	\label{SubSec:ThreeFermion}

	Let us consider different methods of error correction with the color code by considering the full stabilizer group. As we have already mentioned the symmetries we have used to find the matching decoder are connected to the conservation laws of the equivalent system of two decoupled surface codes that can be obtained by unfolding~\cite{Bombin12,Bhagoji15, Kubica15}. Here we look at an alternative unfolding of the color code into two copies of the so-called three-fermion model~\cite{Rowell09, Kesselring18}. We look at the conservation laws of this model and its corresponding symmetries. In addition to fundamental interest, this observation may be valuable to find a decoder for the color code undergoing depolarising noise. Indeed, it is known that the color code undergoing depolarising noise can demonstrate high threshold error rates~\cite{Bombin12, Tuckett19}.
	
	We begin by redefining some of the notation we used above where we only considered a bit-flip error model. In addition to measurement errors let us now assume an error model where Pauli-X, Pauli-Y and Pauli-Z errors can occur. For our discussion, in addition to the Pauli-Z stabilizers $S_f^Z = \prod_{v \in \partial f} Z_v$ we already defined, we also require Pauli-X stabilizers $S_f^X = \prod_{v \in \partial f} X_v$ and Pauli-Y stabilizers $S^Y_v = S^Z_v S^X_v $. We therefore have symmetries
	\begin{equation}
		\Sigma_{\mathbf{c}, P} = \left\{ S_f^{P} \in \Sigma_{\mathbf{c}, P} | \mathbf{col}(f) \not= \mathbf{c}  \right\},
	\end{equation}
	where $P = X,\,Y,\,Z$ is a Pauli label.

	We will now redefine defects $\mathbf{c}_P$ with both a color label $\mathbf{c}$ and a Pauli label $P$ according to the convention in~\cite{Kesselring18}. We say that a face $f$ supports a defect with Pauli label $P$ if $S^P_f = +1$ but all other stabilizers at $f$ give the -1 outcome. The defect also takes the color label corresponding to $\mathbf{col}(f)$. Note that 
	\begin{equation}
		S_f^X S_f^Y S_f^Z = +1, \label{Eqn:Constraint}
	\end{equation}
	by definition. Therefore there must be an even parity of violated stabilizers at any given site. Roughly speaking this convention means a defect has an $X$ label if it is created at the end point of a string of Pauli-X errors. Likewise a $Z$($Y$) label if the defect is created at the end point of a Pauli-Z(Pauli-Y) errors. 
	
	Having now extended the description of defects of the color code to include both a color and a Pauli label, we can also extend the conservation laws among the defects that respect the symmetries. By definition, the stabilizers of $\Sigma_{\mathbf{c},P}$ return $-1$ outcomes at faces that support defects $\mathbf{c'}_{P'}$ such that neither $\mathbf{c'} = \mathbf{c}$ nor $P' = P$ can be true. For instance, the existence of the symmetry $\Sigma_{\mathbf{r},Z}$ indicates that the collection of all defects with labels $\mathbf{g}_X,\mathbf{b}_X,\mathbf{g}_Y,\mathbf{b}_Y$ over the lattice respect parity conservation, up to the lattice boundaries.

\begin{figure}
\includegraphics{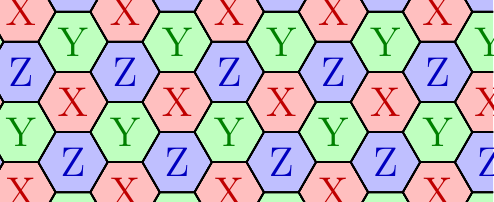}
\caption{\label{Fig:XYZ} The product of Pauli-X stabilizers, Pauli-Y stabilizers and Pauli-Z stabilizers on, respectively, red, green and blue faces gives rise to a symmetry.}
\end{figure}

	In Ref.~\cite{Brown20parallelized} it was proposed we could find alternative symmetries for the color code to obtain variations of the matching decoder, see Appendix~D; example~5. It is pointed out, for instance, that the product of $S_f^X$ stabilizers on red faces, $S^Y_f $ on green faces, and $S^Z_f$ on blue faces gives rise symmetry up to lattice boundaries, see Fig.~\ref{Fig:XYZ}. In general, on a lattice with closed boundary conditions, we have a symmetry 
	\begin{equation}
		\Sigma(\mathbf{c}^X,\mathbf{c}^Y,\mathbf{c}^Z) = \left\{ S_f^P \in \Sigma(\mathbf{c}^X,\mathbf{c}^Y,\mathbf{c}^Z) |\textbf{col}(f) = \mathbf{c}^P \right\},
	\end{equation}
	where $\mathbf{c}^X,\,\mathbf{c}^Y,\,\mathbf{c}^Z$ are colors that all take different values. Locally correctable clusters of defects can be obtained by combining groups of defects that are matched over four different variations of this symmetry where variations are obtained with permutations over the three elements $(\mathbf{c}^X,\mathbf{c}^Y,\mathbf{c}^Z)$. Specifically, the four variations of this matching must include two even permutations of $(\mathbf{c}^X,\mathbf{c}^Y,\mathbf{c}^Z)$ and two odd permutations. We elaborate on this point shortly. 
	
Once again, by definition, we have that a subset of defects over the lattice respect a parity conservation law that corresponds to a given symmetry. For symmetry $\Sigma(\mathbf{c}^X,\mathbf{c}^Y,\mathbf{c}^Z)$, we pair all types of defects on the lattice neglecting only ${\mathbf{c}^X}_X$, ${\mathbf{c}^Y}_Y$ and ${\mathbf{c}^Z}_Z$. Indeed, these three defect types are not detected by any of the stabilizers in $\Sigma(\mathbf{c}^X,\mathbf{c}^Y,\mathbf{c}^Z)$.

We can understand the symmetries of the color code $\Sigma(\mathbf{c}^X,\mathbf{c}^Y,\mathbf{c}^Z)$ in terms of decoupled copies of the three-fermion model~\cite{Kesselring18}. The connection between the color code and the three-fermion model is explained in Ref.~\cite{Kesselring18}; Appendix~B. See also recent work~\cite{Roberts20} on the three-fermion model. Let us give some brief remarks on the model. We label individual charges $\alpha$, $\beta$ and $\gamma$. We can think of the charges of the three-fermion model as different types of defects for the syndrome pattern of some abstract code that satisfies
\begin{equation}
\left |\# {\bf\alpha} \right|_2 = \left|\# \beta \right|_2 = \left|\# \gamma \right|_2. \label{Eqn:FermionicConservation}
\end{equation}
where $| \# f |_2 = 0,\,1$ denotes the number of defects of type $f = \alpha,\,\beta,\,\gamma $ over the global system modulo 2. While any given value $|\# f|_2$ may take an odd or even parity, the sum of the number of two types of fermion must take an even parity as a consequence of Eqn.~(\ref{Eqn:FermionicConservation}), i.e., $|\# f'|_2  + | \#f |_2 = 0$ modulo 2 with $f',\,f = \alpha, \,\beta,\, \gamma$. We therefore obtain a charge parity conservation law that allows us to apply matching to find a correction for a syndrome with this structure. As an aside, we note that these conservation laws reflect those of the color code undergoing a bit-flip error model, see Eqn.~(\ref{Eqn:ConservationLaw}).

In Ref.~\cite{Kesselring18} it is explained that we can express color code defects with labels $\mathbf{c}_P$ presented above in terms of pairs of fermionic charges; one from each of two decoupled layers. We append a positive or negative superscript, $\pm$, to the fermion labels to denote which of the two decoupled layers a given fermion belongs to. We forego the details of the mapping between the three-fermion model and the color code~\cite{Kesselring18} that emerges at the level of the anyonic quasiparticle excitations of the underlying topological phase of the color code~\cite{Bombin06, Rowell09, Bombin12}. For this discussion it is sufficient to state how the color code defects equate with pairs of fermions as shown in Table~\ref{Tab:Fermions}.

\begin{table}
\begin{tabular}{c||c|c|c}
 & $X$& $Y$ &$Z$ \\
 \hline \hline
$\mathbf{r}$  & $\alpha^+ \alpha^- $& $ \beta^+\beta^- $&$ \gamma^+ \gamma^- $\\
 \hline
$\mathbf{g}$ & $\gamma^+\beta^- $& $\alpha^+ \gamma^- $& $ \beta^+\alpha^- $\\
 \hline
 $\mathbf{b}$ & $\beta^+ \gamma^-$&$ \gamma^+ \alpha^- $&$ \alpha^+ \beta^-$ 
 \end{tabular}
 \caption{Color code defects $\mathbf{c}_P$ expressed in terms of pairs of fermions $\alpha^\pm$, $\beta^\pm$ and $\gamma^\pm$ of the three-fermion model. The superscripts $\pm$ denote different copies of the three-fermion model. Colors for defects $\mathbf{c} = \mathbf{r},\, \mathbf{g},\, \mathbf{b}$ vary with rows and Pauli labels $P = X,\, Y,\, Z$ vary along columns of the table.\label{Tab:Fermions}}
\end{table}

By examination of Table~\ref{Tab:Fermions} we find that the defects that are identified by the stabilizers of $\Sigma(\mathbf{c}^X,\mathbf{c}^Y,\mathbf{c}^Z)$ symmetries correspond to the conservation laws of the fermionic theory; Eqn~(\ref{Eqn:FermionicConservation}). Let us take, for example, $\Sigma(\mathbf{r}, \mathbf{g}, \mathbf{b})$ with $\mathbf{c}^X = \mathbf{r}$, $\mathbf{c}^Y = \mathbf{g}$ and $\mathbf{c}^Z = \mathbf{b}$ such that we pair all defects on the lattice with the exception of $\mathbf{r}_X$, $\mathbf{g}_Y$ and $\mathbf{b}_Z$. These three defects are those along the leading diagonal of Table~\ref{Tab:Fermions}. We observe that all other defects include either a $\beta^+$ fermion or a $\gamma^+$ fermion. On the other hand all the defects along the leading diagonal include an $\alpha^+$ fermion rather than a $\beta^+$ or a $\gamma^+$ fermion on the positive copy of the three-fermion model. Therefore, by pairing all defects identified by stabilizers $\Sigma(\mathbf{r}, \mathbf{g}, \mathbf{b})$ we obtain pairs of color code defects that satisfy $|\# \beta^+|_2 + |\# \gamma^+|_2 = 0$ under the three-fermion mapping. With the arguments given above we have that any collection of color code defects is satisfied if, say $|\# \beta^\pm|_2 + |\# \gamma^\pm|_2 = 0$ and $|\# \alpha^\pm|_2 + |\# \beta^\pm|_2 = 0$ are all satisfied on both the positive and negative copies of the three-fermion theory.

Curiously, we find that we pair for the conservation laws of the positive copy of the three-fermion theory if we take the defects measured by symmetries with an even permutation of the three colors in the argument of $\Sigma(\mathbf{r}, \mathbf{g}, \mathbf{b})$, and we pair for the conservation laws of the negative copy of the three-fermion model for odd permutations of $\Sigma(\mathbf{r}, \mathbf{g}, \mathbf{b})$. For instance, $\Sigma(\mathbf{g}, \mathbf{b}, \mathbf{r})$ with $\mathbf{c}^X = \mathbf{g}$, $\mathbf{c}^Y = \mathbf{b}$ and $\mathbf{c}^Z = \mathbf{r}$ we obtain pairs of defects that respect $|\alpha^+|_2 + |\beta^+|_2 = 0$. In contrast we have that pairing the defects measured by the stabilizers of $\Sigma(\mathbf{r}, \mathbf{b}, \mathbf{g})$ gives pairs of defects satisfying $|\beta^-|_2 + |\gamma^-|_2 = 0$.

	\subsection{Fault-tolerant error correction}
	
	Minimum-weight perfect matching decoders can be generically adapted to the fault-tolerant setting where we assume unreliable measurements~\cite{Dennis02, Brown20parallelized}. To identify measurement errors we repeat stabilizer measurements to collect syndrome data over a long time. We then define a defect in spacetime where the stabilizer measurement performed at time $t$, denoted $S(t)$, differs from the outcome of the measurement $S(t-1)$. We therefore recover a defect parity symmetry in spacetime for each subset of stabilizer operators that respects a symmetry. Specifically, up to boundaries, we have a symmetry for checks $S(t)S(t-1)$ over all $t$ and $t-1$ for $S\in\Sigma$ with $\Sigma \subseteq\mathcal{S}$ a symmetry of the stabilizer group. See Appendix D example 8 in Ref.~\cite{Brown20parallelized} for a discussion.

	\begin{figure}
		\includegraphics{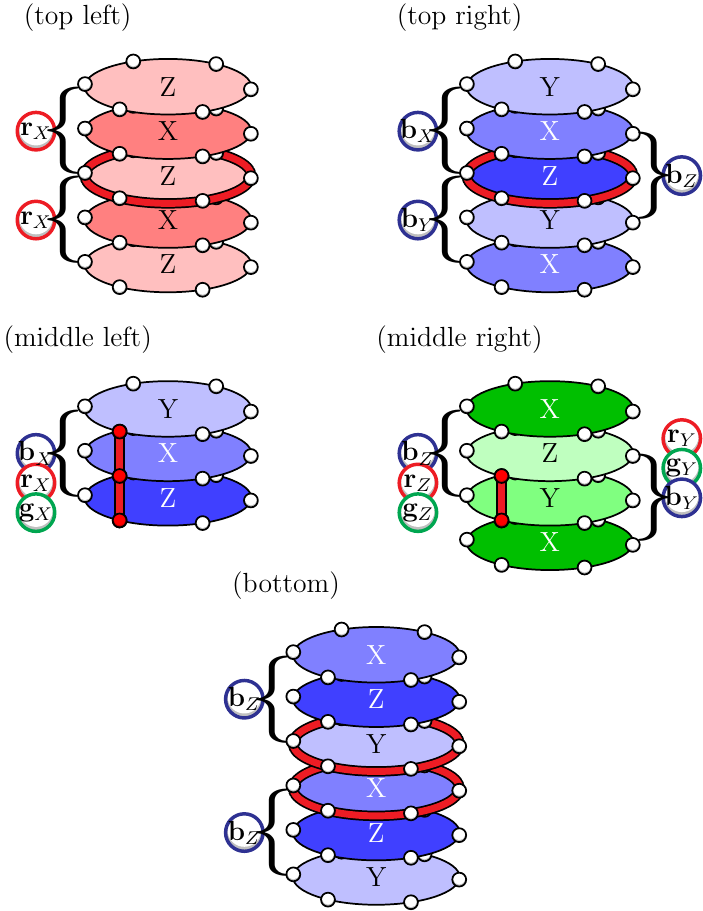}
		\caption{\label{Fig:MeasurementErrors} The syndrome of stabilizers measured in spacetime. In all figures time runs in the upward direction. (top left)~The Pauli-X stabilizer and the Pauli-Z stabilizer of a common plaquette are measured over time in an alternating pattern. If no errors occur, we expect $S_f^Z(T)=+1$ for all measurements $T = t-1,\, t,\,t+1$. A measurement error on the central face occurs, we therefore create a pair of $r_X$ defects detected by checks $S_f^Z(t)S_f^Z(t-1) = -1$ and $S_f^Z(t+1)S_f^Z(t) = -1$. (top right)~We measure a periodic sequence of Pauli-X, Pauli-Y and Pauli-Z stabilizers and define a check as the product of three time-adjacent measurements. A single measurement error creates three defects. (middle left)~A single physical Pauli-X error violates three adjacent checks of different colors. A single blue check is shown. (middle right)~A Pauli-X error occurring between the measurement of Pauli-Y stabilizer and Pauli-Z stabilizer measurements creates six defects. (bottom)~A pair of measurement errors creates two defects of the same type with temporal separation.}
	\end{figure}

	With the extended symmetries defined we show an example of a measurement error in Fig.~\ref{Fig:MeasurementErrors}(top left). The figure shows a Pauli-Z stabilizer and a Pauli-X stabilizer measured sequentially over time on the same face where time runs upwards. A measurement error occurs on the central Pauli-Z measurement. This creates a pair of $\mathbf{r}_X$ defects separated over time given that the $S_f^X$ stabilizer has not been violated. Clearly, the creation of a pair of $\mathbf{r}_X$ defects respects the defect conservation laws of the color code. Moreover, one can check that a single Pauli-X error on a qubit at some time will simultaneously create three defects on a common time plane; one $\mathbf{r}_X$ defect, one $\mathbf{g}_X$ defect and one $\mathbf{b}_X$ defect. We can therefore adapt our methods of decoding proposed above in the spacetime picture, where we pair defects of two of the three colors for a common Pauli label. In this setting we can take a common simplification where we view Pauli-Y errors as creating two triples of defects; one triple of $\mathbf{r}_X,\, \mathbf{g}_X, \,\mathbf{b}_X$ defects and one of $\mathbf{r}_Z,\, \mathbf{g}_Z, \,\mathbf{b}_Z$, both of which can be decoded independently.

	We can use the notion that there are symmetries between different Pauli labels as described in SubSec.~{\ref{SubSec:ThreeFermion}} to propose alternative fault-tolerant error-correction protocols. Our strategy to repeatedly measure stabilizers means we can compare the values of stabilizer measurements taken at different times to test for agreement. This is because, assuming no errors occur, pairs of time adjacent measurements are constrained to agree. In general we can propose using the results of different subsets of stabilizer measurements that are constrained to agree to identify both measurement and physical errors. For instance, it follows from Eqn.~(\ref{Eqn:Constraint}) that the product $S_f^X$, $S_f^Y$ and $S_f^Z$ are constrained to give a constant result. A violation of this constraint indicates that a local error has occurred; either a physical error or a measurement error.

	Assuming a phenomenological noise model where we can choose to measure any one face operator per unit time, we propose measuring Pauli-X, Pauli-Y and Pauli-Z stabilizers sequentially. As we will explain, we find that this improves the distance of the code to time-like correlated errors. Now, given that the product of any three sequential stabilizer measurements are constrained to give the same result, we find that more defects are created for a given measurement error. In Fig.~\ref{Fig:MeasurementErrors}(top right) we show five measurements in this sequence. The figure shows a measurement error occurring on the middle measurement in the sequence. This measurement error will violate three of these constraints. These constraints consist of triples of stabilizers that are grouped by the braces at the side of the measurement sequence.

	As with the conventional case already discussed, the defects of the spacetime syndrome produced for the phenomenological noise model respect the same conservation laws as in the idealised case where all measurements are reliable. As always, we color each defect according to the color of the face on which it is detected. Further, we give each defect a Pauli label according to the middle of the three stabilizer measurements for some given constraint. As shown in Fig.~\ref{Fig:MeasurementErrors}(top right), a single measurement error produces three defects of the same color, where each defect has a different Pauli label. Indeed, we show $\mathbf{b}_X$, $\mathbf{b}_Y$ and $\mathbf{b}_Z$ defects in the figure. This configuration is consistent with the defect conservation laws of the color code discussed in Subsec.~\ref{SubSec:ThreeFermion} for the case of depolarising noise and ideal measurements.

	Let us show why this convention is consistent with that we have proposed above. We will concentrate on a single Pauli-X label, but our discussion is symmetric over any choice of Pauli error. We defined a defect with a Pauli-X label appearing at the end point of a string of Pauli-X errors. In Fig.~\ref{Fig:MeasurementErrors}(middle left) we suppose a Pauli-X error occurs on a single qubit during the interval marked by the red line. As the Pauli-X error is not identified by Pauli-X stabilizer measured in the middle of the displayed sequence, it does not matter if the error occurs before or after this measurement. We find that only one blue defect is produced that we label $\mathbf{b}_X$. The time adjacent constraints are not affected by the Pauli-X error as the error violates an even parity of each of the stabilizer measurements of both checks. Defects of other colors, $\mathbf{r}_X$ and $\mathbf{g}_X$, must also be created but are not shown in the figure.
	
	It remains to check the syndrome produced if a Pauli-X error occurs in between a Pauli-Y and Pauli-Z stabilizer measurements, see Fig.~\ref{Fig:MeasurementErrors}(middle right). In this case, we produce six defects at two different temporal planes. These are $\mathbf{r}_Y$, $\mathbf{r}_Z$, $\mathbf{g}_Y$, $\mathbf{g}_Z$,  $\mathbf{b}_Y$, $\mathbf{b}_Z$. In the figure we find that only the Pauli-Z stabilizer is violated of the two green checks that are shown. Once again, the blue and red checks behave equivalently but are not shown. Given that for each color we have a $\mathbf{c}_Y$ defect and a $\mathbf{c}_Z$ defect, and each of these are locally consistent with a $\mathbf{c}_X$ defect, we can maintain the convention we have proposed for a Pauli-X error occurring at any time on any qubit in this spacetime syndrome.

	The spacetime defect configurations for this alternative procedure for fault-tolerant error correction shares a pleasing symmetry between single-qubit physical errors and single-measurement errors. Indeed, a single qubit physical error creates three defects of alternate colors, but with the same Pauli label, see e.g.~\ref{Fig:ColorCode}(c). In contrast, a single measurement error creates three defects of the same color with three distinct Pauli labels, see Fig.~\ref{Fig:MeasurementErrors}(middle left). Likewise, in this scheme, we require two measurement errors to produce two defects of the same type. In Fig.~\ref{Fig:MeasurementErrors}(bottom) we show an error configuration that produces two $\mathbf{b}_Z$ defects. This shares a commonality with physical errors where we require two bit-flip errors to separate two defects of the same type, see Fig.~\ref{Fig:ColorCode}(d). It may be valuable to find better decoders to exploit this structure. Further development of the theory of symmetries and decoding might lead to such decoding algorithms~\cite{Brown20parallelized}.

	We can obtain better fault-tolerant error-correction procedures by considering alternative stabilizer readout circuits, see for example~\cite{Zeng19, Ashikhmin20, Delfosse20}. Long sequences of measurement errors can lead to logical errors when we perform code deformations~\cite{Vuillot18}. We obtain an improved code distance against time-like logical errors using our new stabilizer readout procedure. For the following discussion we assume we can make any one stabilizer measurement of our choice, $S^X_f$,  $S^Y_f$, or  $S^Z_f$, for each face per unit time. 
	
	With these assumptions, if we consider the standard protocol where we alternately measure Pauli-X stabilizers and Pauli-Z stabilizers, we find that a time-like logical error occurs if half of the measurements experience errors. For example, we obtain a timelike logical error if all of the Pauli-Z stabilizer measurements experience a measurement error. In contrast, an undetectable string-like logical operator using the cyclic protocol requires that two-thirds of the stabilizer measurements experience measurement errors. For instance, measurement errors that occur on all of the Pauli-X measurements and all of the Pauli-Y measurements will lead to an undetectable error. 
	
	Therefore, under this phenomenological noise model, if we aim to reach a distance against time-like logical errors, $d_t$, in the standard protocol we must measure for $2d_t$ rounds of stabilizer measurement but only $3d_t / 2$ rounds of measurement in the alternative cyclic protocol. We therefore complete code deformations in three-quarters of the time using the cyclic readout protocol as compared with the standard protocol, that is $(3 d_t/ 2) / (2d_t) = 3/4 $. This is achieved with a simple local rotation on stabilizer measurements we would otherwise need to perform on a standard implementation of the color code.
	
	We remark that we can obtain similar resource saving for foliated color-code models~\cite{Bolt16, Brown20universal}. It is easy to include Pauli-Y stabilizer measurements in a foliated fault-tolerant scheme using type-II foliated qubits~\cite{Brown20universal}. Interestingly, the resource state we obtain if we produce a foliated system with type-I qubits to measure the standard stabilizer readout pattern~\cite{Bolt16} is the same as that which we obtain if we use type-II foliated qubits to measure the cyclic stabilizer readout pattern. The only difference between the two models is that in the latter case, some of the Pauli-X measurements used to read out the resource are replaced with Pauli-Y measurements. Nevertheless, we find a significantly different syndrome pattern, and a non-trivial improvement in resources. Given the development of better decoders, we can therefore obtain a significant resource overhead with a minor basis rotation to the measurement devices of the hardware that is designed to produce such a system.

	To move forward, it will be valuable to evaluate the performance of these error-correction procedures with numerical simulations. This will require finding stabilizer readout circuits to test different fault-tolerant protocols undergoing circuit noise. Stabilizer readout circuits to measure the sequence of stabilizers in Fig.~\ref{Fig:MeasurementErrors}(top left) have been considered in Refs.~\cite{Stephens14, Chamberland20, Beverland21}. It may be interesting to look for better readout circuits that also include Pauli-Y stabilizer measurements. See also Refs.~\cite{Andrist11, Katzgraber13, Andrist16} where remarkably high fault-tolerant thresholds are found using statistical-mechanical methods. It may also be interesting to repeat these analyses for alternative stabilizer readout protocols.

	\subsection{Higher-dimensional color codes}
	
	Let us look at how the methods for decoding we have introduced generalise to higher-dimensional color codes. See Refs.~\cite{Bombin07, Bombin07a, Bombin13, Kubica15a} for a detailed definition where these codes are introduced and Refs.~\cite{Brown16a, Turner20, Beverland21} where decoders for the three-dimensional color code are implemented. We will focus on the example of decoding point-like defects for the three-dimensional color code, but we expect that the principles we are developing for decoding point-like defects are general to decoding problems. We leave these generalisations as an exercise for the reader. We write the dimensionality of the system as $D$ where a general statement is given although we largely focus on the case where $D=3$.

	We can define a $D$-dimensional color code with qubits on the vertices of a $(D+1)$-valent lattice with $(D+1)$-colourable cells. Pauli-X stabilizers are assigned to the $D$-dimensional cells of the lattice. We focus on error correction with these stabilizers that identify Pauli-Z errors. Boundaries can also be defined with one of $D+1$ colours assigned such that no cell of the assigned color is found at the boundary. The $D$-dimensional color code defined on a $D$-dimensional tetrahedral lattice with one boundary of each colour encodes a single qubit.

	Let us motivate the generalisation of the unified lattice matching decoder by looking at how the challenges of decoding with a restricted lattice generalise with dimension. These considerations will be important when comparing the resource cost of quantum computation using color codes of different dimensionality. We can decode point defects of the $D$-dimensional color code using a restricted lattice. Up to the lattice boundaries, the product of all of the cells of two different colors, independent of $D$, give the identity operator. We therefore obtain a symmetry that can be used to obtain a restricted lattice for decoding~\cite{Kubica19}. The result of matching the defects on $D$ of these sublattices can be combined to find a correction. 
	
	\begin{figure}
		\includegraphics{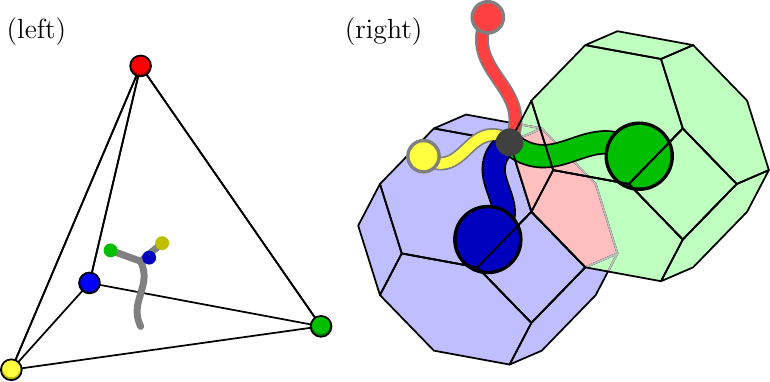}
		\caption{\label{Fig:3Dproblems} Decoding point defects for the color code in higher dimensions. (left)~An error of weight $\sim d/4$ on the three-dimensional tetrahedral color code that may lead a matching decoder on a restricted lattice to a logical failure. The string-like error stretches up from the red boundary to the centre of the lattice such that the three defects can be paired to the boundary of their corresponding colour with $ \lesssim d / (D+1)$ errors. (right)~An error on a single qubit and its syndrome on the restricted lattice of blue and green cells. A single error on any of the six qubits of the face separating the two cells, shaded in pink, will give rise to the same syndrome with respect to the restricted lattice.}
	\end{figure}
	
	In SubSec.~\ref{SubSec:HighWeightErrors} we identified an issue with decoding the color code on a two-dimensional tetrahedral lattice (a triangle) using separate restricted lattices, namely, that errors of weight $\sim d/3$ may lead to logical failure. Ideally we should aim to design decoders that can correct up to $d/2$ errors. In fact, as dimensionality increases, we expect errors of lower weight to lead to logical failure. Indeed, a string of length $O(d / (D+1))$ that stretches from a boundary towards the centre of a $D$-dimensional tetrahedral lattice may cause a restricted lattice decoder to cause a failure. In Fig.~\ref{Fig:3Dproblems}(left) we show one such error of weight $\sim d/4$ for the three-dimensional color code. Roughly, we might expect to obtain a quadratic improvement in logical failure rate for the three-dimensional color code if we can produce a decoder that can tolerate errors of weight up to $(d-1)/2$.
	
	We also find that the issue of syndrome degeneracy on a restricted lattice increases with dimensionality. In Fig.~\ref{Fig:ColorCodeSymmetry}(d) we show an error with red defects that are neglected by the restricted lattice. See also Fig.~\ref{Fig:RestrictedLattice}. Nevertheless, one should expect to be able to use this information to obtain a more accurate correction. In SubSec.~\ref{SubSec:Degeneracy} we argued that decoding on the unified lattice gives us access to this information, and we give an example where an error of this type is corrected by a matching decoder on the unified lattice that might not otherwise be corrected by matching on separate restricted lattices.
	
	In two dimensions, a bit flip on one of the two qubits separating two cells can give rise to the same syndrome on the restricted lattice. Therefore, there may be up to $2^w$ different error configurations that can give rise to the same syndrome on a restricted lattice for an error of weight $w$ since any single qubit error can be moved onto the opposite qubit of its edge without changing the syndrome. We remark however that $2^w$ is an upper bound to account for error configurations where both qubits on some given edge separating two cells of the restricted lattice are flipped, in which case a rearrangement can not be made for any such given edge. At low error rates this is quite uncommon.
	
	As dimensionality increases, we find that the degeneracy of the syndrome on a restricted lattice also increases. In Fig.~\ref{Fig:3Dproblems}(right) we show a single error together with its syndrome on two cells of the restricted lattice. In fact, an error on any of the qubits on the face at the intersection of the two cells will give rise to the same restricted syndrome. We therefore find that syndrome degeneracy on the restricted lattice increases as we progress from decoding in two to three dimensions. In three dimensions we have four-colourable lattices that are composed of faces of weight four or six. In addition to the lattice shown in the figure, see for instance~\cite{Brown16a}.

	\begin{figure}
		\includegraphics{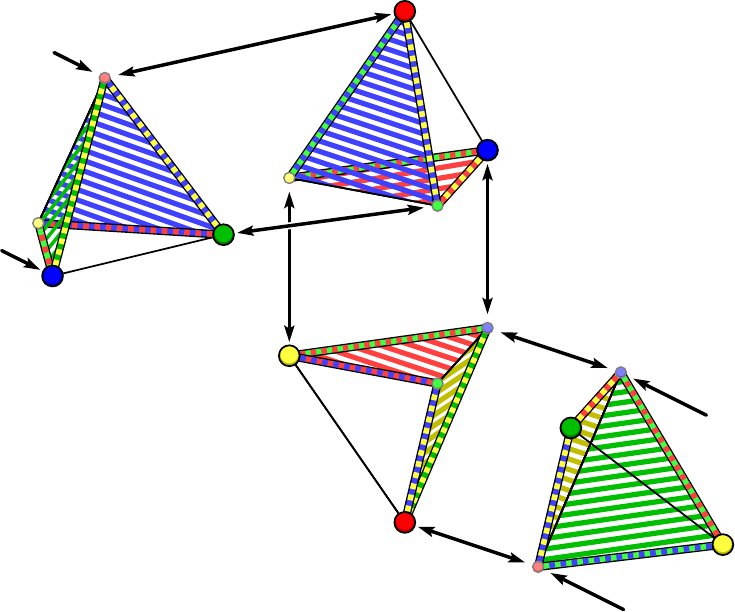}
		\caption{\label{Fig:3Dmobius} Four boundary operators obtained by taking the product of all the cell operators of two of the four colour subsets of stabilizers. The left most tetrahedron shows the product of the blue and the green cell operators. The non-trivial support of the boundary operator lies on the blue and green boundaries, with the exception of the qubits that lie on both the blue and green boundaries. The figure also shows the boundary operator for the red-blue restricted lattice, the yellow-red restricted lattice and the green-yellow restricted lattice. The restricted lattices can be combined by their shared boundaries via the displayed arrows to produce a unified lattice to decode the point excitations of the three-dimensional color code.}
	\end{figure}
	
	The observations above motivate generalising the unified lattice to higher-dimensional systems. Let us concentrate on a three-dimensional color code. In three dimensions we use colors $\mathcal{C} = \left\{ \mathbf{r},\, \mathbf{g} ,\, \mathbf{y} ,\, \mathbf{b} \right\}$. We show a construction for a unified lattice in Fig.~\ref{Fig:3Dmobius}. The figure depicts four boundary operators 
	\begin{equation}b_\mathbf{uv} = \prod_{\mathbf{col}(c)=\mathbf{u},\mathbf{v}}S_c^X,
	\end{equation} 
	where $\mathbf{u} ,\mathbf{v} \in \mathcal{C}$ and $\mathbf{u} \not= \mathbf{v}$. For instance, at the top left of the figure we show $b_{\mathbf{gb}}$ where the support of the boundary operator on the tetrahedron is colored. The operator is supported on the green boundary as qubits on the green boundary support a blue cell but not a green cell. Likewise the blue boundary supports green cells but not blue cells, and therefore the boundary operator has non-trivial support on these qubits. As such the green and blue boundaries are colored. The boundary operator has support on all the edges of the tetrahedron with the exception of the edge connecting the blue and green corners of the lattice, and the edge connecting the red and yellow corners of the lattice. Finally, the boundary operator $b_\mathbf{gb}$ also has support on the blue and green corners of the lattice where only a single blue or green cell operator lie, respectively.

	We find that we can produce a unified lattice by combining four restricted lattices. For instance, the product of four boundary operators; $b_\mathbf{bg}$, $b_\mathbf{rb}$, $b_\mathbf{yr}$, $b_\mathbf{gy}$ returns identity. We can therefore combine their respective restricted lattices to give a unified lattice that we might expect to correct errors of weight greater than $O(d/4)$. We show how these operators are combined in Fig.~\ref{Fig:3Dmobius} giving a generalisation of the M\"obius strip we have studied for the two-dimensional case.
	
	The unified lattice shown in Fig.~\ref{Fig:3Dmobius} is in some sense unsatisfying, since we have six different restricted lattices for the three-dimensional code, and yet we only use four of them to obtain the lattice. We may expect a decoder that incorporates all of the available restricted lattices to perform better, as it captures all of the local correlations between pairs of syndrome defects. However, we do not expect to find any such lattice only using each restricted lattice once. In general we can check that the product of all boundary operators for a $D$-dimensional color code with odd $D$ does not give a symmetry. For odd $D$ we have that 
	\begin{equation}
		\prod b_{\mathbf{uv}} = \prod_c {{S_c^X}}^D = \prod_c {{S_c^X}} \not= \openone.
	\end{equation} 
	where we take the product over all $  (D+1) ! / [ 2! (D-1)! ] = D(D+1) / 2$ boundary operators on the left-hand side of the equation. In three-dimensions this operator has support on the boundaries and the corners of the tetrahedral lattice where the qubits support an odd number of cell operators.
	
	It may therefore be worthwhile considering decoding strategies using alternative unified lattices in odd dimensional color codes. One can conceive of unified lattices that use all of the restricted lattices twice. We also remark that the defects on all of the cells of all $D+1$ colors of an odd $D$-dimensional color code respect a symmetry, up to the lattice boundaries. We may therefore consider using this bulk symmetry together with each of the $ D(D+1)/2$ restricted lattices to produce a unified lattice for $D$-dimensional color codes with $D$ odd. In contrast to the odd dimensional case, for $D$-dimensional color codes with even $D$ we have that $\prod b_{\mathbf{uv}} =  \prod_c  {S_c^X}^D  = \openone$. We can therefore expect to find a consistent generalisation of the unified lattice we have presented in two dimensions to decode point defects for any even-dimensional color-code lattice that combines each of the restricted lattices of the code once and only once. Let us finally say that the arguments we have given for more general unified lattices only indicate that they can exist. We have not made any suggestions for how the restricted lattices should be combined to produce a decoder that performs well. We leave this as an open research problem for the reader.

	\subsection{Single-shot error correction}

	Let us now elaborate on single-shot error correction~\cite{Bombin14a} for the gauge color code~\cite{Bombin15} in terms of symmetries~\cite{Bombin14a, Bombin18transversal, Brown20parallelized}. Single-shot error correction is a procedure to accurately correct errors that have occurred on a code, even if error-detection measurements are unreliable. See numerical implementations of single-shot fault-tolerant decoding algorithms for the gauge color code in Refs.~\cite{Brown16a, Beverland21} and related work using the three-dimensional subsystem toric code in~\cite{Kubica21}. Our exposition will explain how to implement syndrome estimation using minimum-weight perfect matching.
	
	We will give a brief description of the necessary details of the model but we advise the reader to learn about the models we consider for the following discussion in Refs.~\cite{Bombin14a, Bombin18transversal}. The gauge color code is a subsystem code where we measure face operators to infer the values of the cell operators $S^X_c$ of the three-dimensional color code. The gauge color code is self dual such that an equivalent discussion will hold for Pauli-Z face terms to infer the values of cell stabilizers that identify bit-flips. Given we can correct Pauli-X and Pauli-Z errors separately we concentrate our discussion on just Pauli-X face operators $G_f$ and note an equivalent discussion will hold for error correction in the alternate basis.

	Face operators lie at the intersection of pairs of adjacent cell operators. We say that face $f$ has color $\mathbf{uv}$ if neither of its adjacent face operators are color $\mathbf{u}$ or $\mathbf{v}$, and recall that in three dimensions we use four colors. Of course, due to the four-colorability of the color code lattice the two cells separated by the face have different colors. The ordering of the pair of colors of $f$ is not important, i.e. $\textbf{col}(f) = \mathbf{uv} = \mathbf{vu}$. One additional fact we need is that the faces about one cell of the color code are three colorable. A cell of color $\mathbf{x}$ has face operators on its boundary of color $\mathbf{uv}$, $\mathbf{vw}$, and $\mathbf{uw}$ such that each qubit on the boundary of cell $c$ touch exactly one face of each color. These facts mean that we can express cell operators as follows 
	\begin{equation}
		S_c^X = \prod_{\substack{f \in \partial c \\ \textbf{col}(f) = \mathbf{uv}}} G_f, \label{Eqn:StabilizerWithFaces}
	\end{equation}
	where $\mathbf{col}(c) \not= \mathbf{u},\mathbf{v}$ and here, the set $\partial c$ denotes the set of faces on the boundary of $c$. 
	Importantly, the face operators commute with all of the cell operators of the gauge color code such that we can measure them while remaining in an eigenspace of the cell operators. Moreover, as we have that the cell operators are obtained as the product of face operators, we can infer the values of the cell operators using the outcomes of the face measurements.

\begin{figure}
		\includegraphics{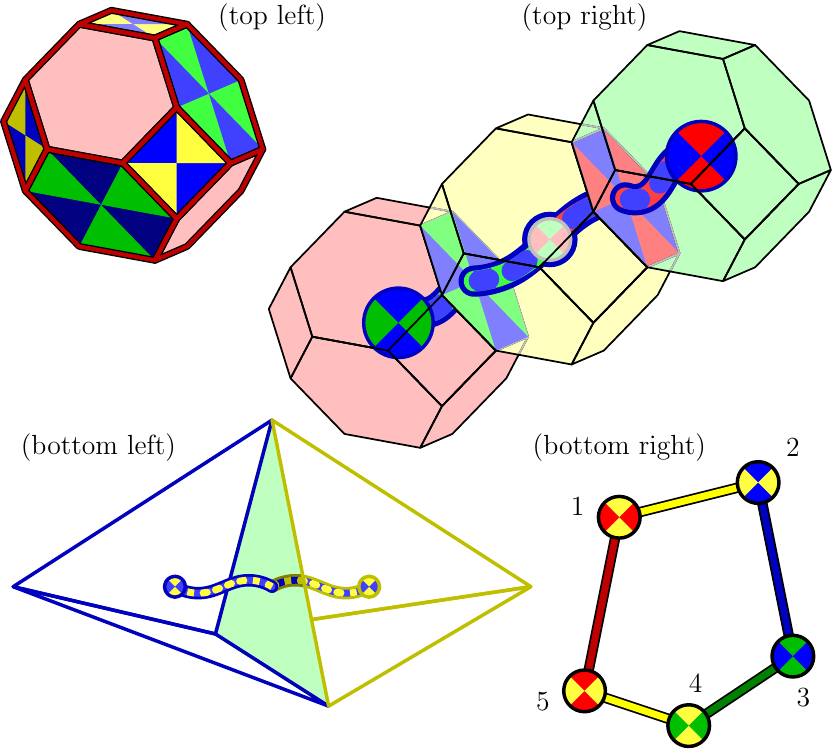}
		\caption{\label{Fig:SingleShot} Single-shot error correction with the gauge color code. (top left)~A blue gauge check. The check is the product of all the face operators about a red cell that include a blue color element. The value of the gauge check is independent of the value of the syndrome of the color code stabilizers. It depends only on measurement errors of its faces. (top right)~Two blue gauge checks are violated on the red and green cell. The blue gauge check of the central yellow cell is not violated. We can consistently pair the blue gauge defects as we necessarily have an even number of them due to their bulk symmetry. (bottom right)~A collection of gauge defects that can be locally corrected. The edges show the results from matching for each of the color charge symmetries. (bottom left)~Matching subroutines for different charge colors are performed in unison. The two constituent charges for a single $\mathbf{yb}$ gauge defect are paired via the green boundary.}
	\end{figure}

	We assume an error model where, in addition to the dephasing errors that occur, a small fraction of face measurements return the value that is opposite to their measurement outcome. Single-shot decoders typically follow a two-step process where we first estimate the error syndrome among errors in the measurement pattern before trying to determine the qubits that have experienced errors. Here we elaborate on the syndrome estimation step. We write down gauge checks, see Fig.~\ref{Fig:SingleShot}(top left), and discuss their symmetries to show how we can perform syndrome estimation using minimum-weight perfect matching on a restricted lattice. We pay specific attention to how the symmetries of the gauge checks are changed at the lattice boundary.
	
	We consider gauge checks that are the product of faces that take two of the three colors at the boundary of each of the cells. We separate these gauge checks $C^\mathbf{u}_c$ into four colors $\mathbf{u} \in \mathcal{C}$ whereby one such gauge check is the product of all the faces that contain color $\mathbf{u}$, i.e., faces $f \in \partial c$ with $\mathbf{col}(f) = \mathbf{uv}$ for any $\mathbf{v}\in \mathcal{C}$. We write this explicitly as
	\begin{equation}
		C_c^\mathbf{u} = \prod_{\substack{f \in \partial c,\, \mathbf{v} \in \mathcal{C} \\ \textbf{col}(f) = \mathbf{uv}}} G_f. \label{Eqn:GaugeCheck}
	\end{equation}
	We show a blue gauge check $C^\mathbf{b}_c$ on a red cell in Fig.~\ref{Fig:SingleShot}(top left). We note that there are no gauge checks $C^\mathbf{u}_c$ on cells of color $\mathbf{u}$ since, by definition, cells of color $\mathbf{u}$ have no faces of color $\mathbf{col}(f) = \mathbf{uv}$ for any value of $\mathbf{v}$. Up to boundaries we have four different symmetries; one associated to each color. We also point out that the values of $C_c^\mathbf{u}$ are independent of the values of stabilizer operators $S_c^X$ since, assuming no measurement errors, we have that $C_c^\mathbf{u} = S^X_c\times S^X_c$ which gives the trivial element of the stabilizer group. As such, these checks strictly identify errors among face measurement outcomes.
	
	It will be helpful to introduce some terminology. Let us call a single check that returns a $-1$ outcome a charge. Charges carry a single color $\mathbf{u}$ that corresponds to the color of the gauge check $C_c^\mathbf{u} = -1$. Up to boundaries, the color charges of a common color respect a global symmetry.
	We also call a pair of color charges at a common cell a gauge defect. A gauge defect carries a color pair that corresponds to the color of its two constituent color charges. A cell $c$ supports a $\mathbf{uv}$ gauge defect if $C_c^\mathbf{u} = C_c^\mathbf{v} = -1$.  We remark that each cell respects a local symmetry that guarantees that there must be an even number of color charges at cell $c$, namely
	\begin{equation}
		C^\mathbf{u}_c C^\mathbf{v}_c C^\mathbf{w}_c = \prod_{f\in \partial c} G_f^2= 1,
	\end{equation}
	with $\mathbf{u},\, \mathbf{v},\, \mathbf{w}$ all distinct colors that are also different from the color of $c$.

	In the bulk of the lattice, a single measurement error on some given face $G_f$ with $\mathbf{col}(f) = \mathbf{uv}$ creates gauge defects of color $\mathbf{uv}$ on its two adjacent cells $c'$ and $c$ with $f \in \partial c' ,\, \partial c$. Specifically we have a syndrome where $C_c^{\mathbf{u}} = C_c^{\mathbf{v}} = C_{c'}^{\mathbf{u}} = C_{c'}^{\mathbf{v}} = -1$, and all other gauge checks give values $+1$. Color charges of some color can be separated further with additional measurement errors on additional faces of the appropriate color. We can regard these measurement errors on faces as strings on the dual lattice. In Fig.~\ref{Fig:SingleShot}(top right) we show two measurement errors on the highlighted faces; one face of color $\mathbf{rb}$ and another of color $\mathbf{gb}$. These give rise to two violated blue gauge checks; one on the red cell and another on the green cell. In the bulk we can consistently match the blue charges as we necessarily have an even number of them.

	In general we can find collections of gauge defects that can be locally corrected provided the collection respects all of the symmetries among the color charges. We can find collections that respect the color charge symmetries by matching for each of the four sets of differently colored charges separately. The edges returned from each matching produces networks of gauge defects such as that shown in Fig.~\ref{Fig:SingleShot}(bottom right). These networks must respect the color charge symmetries since each edge of the network is incident to a pair of charges of a given color.

	Let us now discuss how we can find a correction for these gauge defects. We note that the gauge defects have exactly two incident edges and as such the defects form a one-dimensional chain. We can therefore assign a number to each gauge defect in order such that the $j$-th defect is adjacent to the $j-1$-th and $j+1$-th defect, and the final defect in the chain is adjacent to the first defect. The location of the first defect is chosen arbitrarily. We can then combine pairs of defects in order to find a correction. We first combine the first and second defect. We suppose they have colors $\mathbf{uv}$ and $\mathbf{uw}$ since they must be connected by an edge of color $\mathbf{u}$. The color $\mathbf{v}$ may or may not be different from the charge $\mathbf{w}$. We find a correction to combine these two defects. This removes the first defect, and changes the second defect to give $\mathbf{vw}$. Given that we have removed a pair of $\mathbf{u}$ charges, the network must still respect all of the global color charge symmetries. We then go on to combine the new $\mathbf{vw}$ defect at the second site in the chain and combine it with the third defect. In the case that $\mathbf{v} = \mathbf{w}$ both gauge defects are corrected and we progress to combine the third defect with the next. This process is sequentially repeated along the chain until all of the gauge defects have been removed.

	We remark that in fact we only need to perform matching on three of the four color charge symmetries. Indeed, a network of defects that respects three color charge symmetries necessarily respects the fourth. Suppose the network is supported on cells $c\in R$. One can check that 
	\begin{equation}
		\prod_{c\in R}C_c^{\mathbf{r}} \prod_{c\in R}C_c^{\mathbf{g}} \prod_{c\in R}C_c^{\mathbf{b}} = \prod_{c\in R}C_c^{\mathbf{y}}, \label{Eqn:Dependence}
	\end{equation}
	To see why let us check these cells separately. For a cell that is not yellow, for instance red, we have that $C_c^\mathbf{g} C_c^\mathbf{b} = C_c^\mathbf{y}$. On yellow cells we have that $  C_c^\mathbf{r} C_c^\mathbf{g} C_c^\mathbf{b} =1$, thus giving our result above. Given Eqn.~(\ref{Eqn:Dependence}) holds for a network of defects, and assuming that $\prod_{c\in R}C_c^{\mathbf{u}} = +1$ for $\mathbf{u} = \mathbf{r},\mathbf{g},\mathbf{b}$, then it follows from Eqn.~(\ref{Eqn:Dependence}) that $\prod_{c\in R}C_c^{\mathbf{y}} = +1$ must also hold. This result may be of practical benefit  as it may allow us to choose a preferred set of three matching results to find a more favourable subset of edges.

Let us finally discuss how gauge defects emerge at the boundary of the color code. At the boundary, we can produce single gauge defects that do not respect the global color charge conservation laws. Specifically, we can produce a single $\mathbf{uv}$ defect at a boundary of color that is neither $\mathbf{u}$ nor $\mathbf{v}$. We therefore need to modify the matching problem to account for this at the boundary. Much like the examples we have already considered, we find that we can match differently colored charges via a boundary. For instance, we might consider pairing two differently colored charges, $\mathbf{u}$ and $\mathbf{v}$ via a boundary from which a single $\mathbf{uv}$ gauge defect can be produced, see Fig.~\ref{Fig:SingleShot}(bottom left) where a blue charge and a yellow charge are paired via the green boundary. In general, we should consider matching all of the color charges in unison in a common matching problem where color charges of different colors, say $\mathbf{u}$ and $\mathbf{v}$, can be matched via a path that crosses a boundary of color other than $\mathbf{u}$ or $\mathbf{v}$. We therefore find a generalisation of the unified lattice to the case of matching gauge defects of the gauge color code.
	
  \bibliographystyle{unsrtnat}
   \bibliography{decoders}

\end{document}